\documentclass[prb,twocolumn,showpacs,amsmath,amssymb,superscriptaddress]{revtex4-2}
\usepackage{mathtools}
\usepackage{bm} 
\usepackage{dcolumn} 
\usepackage{graphicx} 
\usepackage{amsmath} 
\usepackage{amssymb} 
\usepackage{amsfonts} 
\usepackage{placeins}

\begin{document}
\title{Anisotropic Fermionic Quasiparticles}

\author{Seongjin Ahn}
\affiliation{Condensed Matter Theory Center and Joint Quantum Institute, Department of Physics, University of Maryland, College Park, Maryland 20742-4111, USA}
\author{S. Das Sarma}
\affiliation{Condensed Matter Theory Center and Joint Quantum Institute, Department of Physics, University of Maryland, College Park, Maryland 20742-4111, USA}

\date{\today}

\begin{abstract}
We have carried out a comprehensive investigation of the quasiparticle properties of a two-dimensional electron gas, interacting via the long-range Coulomb interaction, in the presence of bare mass anisotropy (i.e., with an elliptic noninteracting Fermi surface) by calculating the self-energy, the spectral function, the scattering rate, and the effective mass within the leading order dynamical self-energy approximation. Our theory is exact in the high-density limit. We find novel anisotropic features of quasiparticle properties that are not captured by the commonly used isotropic approximation where the anisotropic effective mass is replaced by the isotropic averaged density-of-states mass. Some of these interesting results are as follows: (1) The many-body renormalization of the quasiparticle spectrum becomes highly anisotropic as the quasiparticle energy increases away from the Fermi energy; (2) the interaction-induced inelastic scattering rate features a strong anisotropy, exhibiting an abrupt jump at different injected energies depending on the momentum direction of the injected electron; (3) the effective mass enhancement is larger (smaller) for the light (heavy) mass, showing that the anisotropy is reduced by interactions.
Our results and analysis show that the unjustified neglect of the mass anisotropy can lead to an incorrect description of quasiparticle properties of the anisotropic system and inaccurate estimates of physical quantities of interest although the use of an equivalent isotropic approximation using the density-of-states effective mass, as is commonly and uncritically performed in the literature, works as a reasonable approximation in many situations. In addition to the complete random phase approximation theory for the anisotropic quasiparticles, we also provide a theory using the simpler plasmon-pole approximation, commenting on its validity for anisotropic self-energy calculations. We comment also on the interaction effect on the Fermi surface topology, finding that the elliptic shape of the bare Fermi surface is preserved, with suppressed ellipticity, in the interacting system to a high degree of accuracy except in the very strongly interacting limit (and for very high bare mass anisotropy). Our theory provides a complete generalization of the existing isotropic many-body theory of interacting electrons to the corresponding anisotropic systems.
\end{abstract}

\maketitle
\section{Introduction}
\begin{figure}[tb]
    \centering
    \includegraphics[width=1\linewidth]{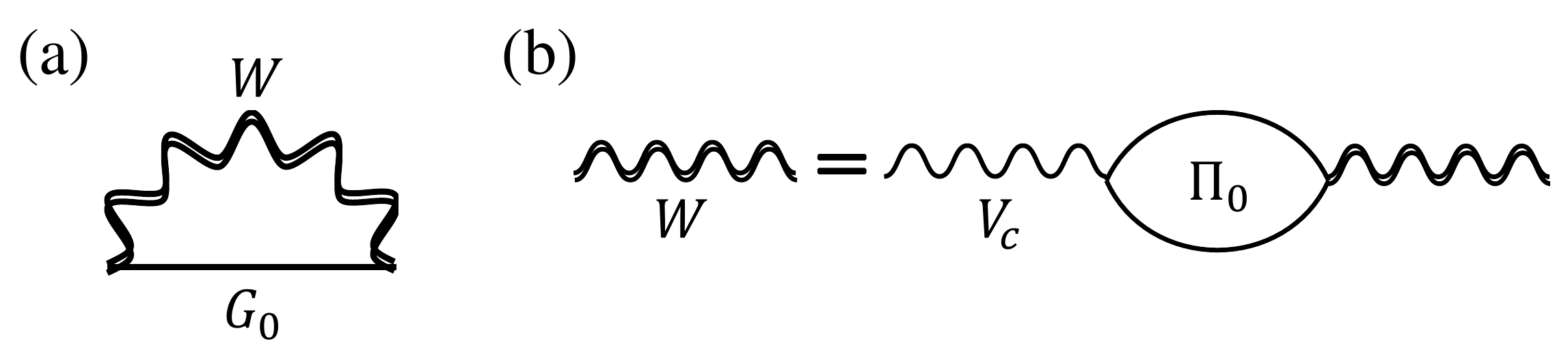}
    \caption{(a) Feynman diagram for the self-energy within the leading order dynamical self-energy approximation. The straight line represents the noninteracting Green's function, and the double wiggly line represents the dynamically screened Coulomb interaction within the RPA, which is obtained by summing up an infinite series of bubble diagrams as shown in (b).    }
    \label{fig:GW_diagram}
\end{figure}

Since the development of Landau's Fermi liquid theory, the quasiparticle concept has been the most popular approach to many-body problems in interacting Fermi systems because of its wide success in describing the low-energy excitations of metals and semiconductors. This is true for both short-range interactions as in normal liquid $^3$He and long-range Coulomb interactions as in solid state systems such as metals and semiconductors-- the current paper is on two-dimensional (2D) electrons interacting via the long range Coulomb interaction. The advantage of Landau's Fermi liquid theory is that it allows us to continue to use the independent particle scheme by reducing the many-body problem into an effective single particle problem by establishing a one-to-one correspondence between the eigenstates of the interacting system and those of the effective noninteracting single particle system. The continued existence of the Fermi surface, with a T=0 discontinuity in the momentum distribution function at the Fermi wave number, is the key feature of the Landau theory, leading to the existence of weakly interacting quasiparticles acting essentially like the noninteracting electrons, except with renormalized parameters, such as effective mass, etc. In the weakly interacting limit, 
the renormalized single particle properties can be exactly calculated using the diagrammatic perturbative many-body theory involving an expansion in the dynamically screened interaction keeping only the bubble or ring diagrams (see Fig.~\ref{fig:GW_diagram}) \cite{quinn1958electron,abrikosov2012methods}, which has been extremely successful in describing quasiparticle excitations in regular materials. In particular, this leading-order self-energy theory, which is exact in the small $r_s$ limit, keeps the infinite series of the random phase approximation (RPA) diagrams in the screened interaction which are the most divergent terms in the high-density limit, thus accounting for the infra-red divergence of the long-range Coulomb coupling correctly. Interestingly, this leading-order self-energy theory, although exact only in the $r_s < 1$ limit, is known to describe well the many-body renormalization effects in metals with $r_s\sim2-6$.  Therefore, empirically the theory appears to be valid in the strongly interacting $r_s>1$ regime also.
In any case, our goal is to develop the many body theory for the anisotropic 2D electron gas interacting via the long-range Coulomb interaction at the same level of sophistication as that existing for the corresponding isotropic system \cite{quinn1958electron, abrikosov2012methods, Rice1965, Lundqvist1967a, Lundqvist1967}.

The quasiparticle properties of an ideal two- and three-dimensional electron gas with an isotropic parabolic energy dispersion have been widely studied in previous literature \cite{quinn1958electron,  Rice1965, Lundqvist1967a, Lundqvist1967, Jalabert1989, Beni1978,fetter2012quantum,mahan2000many, abrikosov2012methods, Lundqvist1970} using the leading order RPA self-energy approach. However, there has never been a study of the quasiparticle properties of an electron gas in the presence of mass anisotropy characterized by an anisotropic energy dispersion given by 
\begin{equation}\label{eq:noninteracting_energy}
    \varepsilon_{\bm k}= \frac{k_x^2}{2m_\mathrm{H}} + \frac{k_y^2}{2m_\mathrm{L}},
\end{equation}
where $m_\mathrm{H}$ and $m_\mathrm{L}$ denote the heavy and light masses, respectively. The bare Fermi surface here is elliptical, and the system no longer has the spherical symmetry of isotropic systems. In fact, most of theoretical studies have neglected the mass anisotropy by employing the isotropic approximation where one replaces the anisotropic effective mass with the isotropic averaged density-of-states mass defined as $m_\mathrm{DOS}=\sqrt{m_\mathrm{H}m_\mathrm{L}}$ \cite{Luttinger1955, Brinkman1972,  Combescot1972, Beni1978}. This is sharp in contrast to the fact that mass anisotropy is common in many electronic materials such as silicon and germanium \cite{Ando1982}. Such spherical isotropic approximation using the density-of-states mass, although used extensively in the literature for simplifying many-body calculations involving anisotropic materials, has never been rigorously theoretically justified in the existing literature despite its wide uncritical use. Anisotropy not only just breaks the spherical symmetry (complicating the theory considerably), but also introduces two distinct many-body interaction parameters corresponding to the dimensionless Coulomb coupling parameter $r_s$ now being different for $m_\mathrm{H}$ and $m_\mathrm{L}$, with $r_s^\mathrm{H} > r_s^\mathrm{L}$.  This is simply because the Bohr radius ($\sim1/m$) now depends on which effective mass (heavy or light) is coming into its definition.

The objective of this paper is to provide a comprehensive study of the bare mass anisotropy effects on quasiparticle properties of two dimensional electron liquids by investigating the self-energy, the spectral function, the inelastic scattering rate, the exchange-correlation potential, and the effective mass within the RPA \cite{Rice1965, Lundqvist1967, Lundqvist1967a, abrikosov2012methods, quinn1958electron} as shown in Fig.~\ref{fig:GW_diagram}. Our results and analysis reveal many unique anisotropic features originating from the imbalance between the two bare masses, which cannot be captured by the isotropic approximation using the density-of-states mass. Our self-energy theory maintains the full mass anisotropy within the leading order ring-diagram dynamical screening approximation, which is exact in the high-density limit. For comprehensiveness, we also discuss the validity of the simpler anisotropic plasmon-pole approximation (PPA) for the anisotropic system by explicitly comparing the RPA results with the corresponding PPA results \cite{Lundqvist1967a, Lundqvist1967}.

This paper is organized as follows. In Sec.~\ref{sec:GW} we introduce the formalism of the self-energy calculation for the Coulomb interaction within RPA and the PPA. In Sec.~\ref{sec:self_energy} we present and analyze results for the anisotropic self-energy and spectral function, and compare them with the isotropic results obtained using the isotropic density-of-states mass. From the calculated dynamical self-energies we present the inelastic scattering rates in Section~\ref{sec:scattering_rate}, discussing a peculiar anisotropic feature that has no isotropic analog. In Sec.~\ref{sec:effective_mass} we present our results for renormalized effective masses along the heavy- and light-mass directions for a wide range of $r_s$, demonstrating that the anisotropy of the system is suppressed by interactions. 
We define the dimensionless interaction parameter $r_s$ in the usual manner by taking it to be the average inter-particle separation measured in the units of Bohr radius: $r_s= me^2/\hbar^2 (\pi n)^{1/2}$, where $m=m_\mathrm{DOS}=(m_\mathrm{H} m_\mathrm{L}) ^{1/2}$ is the bare density-of-states effective mass, and $n$ is the electron density.
Section~\ref{sec:conclusion} contains a summary and conclusions. The two Appendices \ref{sec:appendixA} and \ref{sec:appendixB} respectively present results for the Fermi surface topology change (which is extremely small, $\sim0.1\%$, in general) and for self-energies for additional values of $r_s$ and $m_\mathrm{H}/m_\mathrm{L}$ ratios not presented in the main text.

\section{Theory} \label{sec:GW}

\subsection{RPA Theory}
The self-energy at zero temperature within the RPA scheme is given by [Fig.~\ref{fig:GW_diagram} (a)]
\begin{equation}
    \Sigma({\bm k},\omega)\!=-\!\int\!\frac{d^2 q}{(2\pi)^2}
    \!\int\!\frac{d \nu}{2\pi i}
    W(\bm q,\nu) 
    G_0(\bm k + \bm q, \nu+\omega),
    \label{eq:GW_selfenergy}
\end{equation}
where 
\begin{equation}
 G_0(\bm k, \omega)=\frac{1-n_\mathrm{F}(\xi_{\bm k})}{\omega-\xi_{\bm k }+i\eta}
 +
 \frac{n_\mathrm{F}(\xi_{\bm k})}{\omega-\xi_{\bm k }-i\eta}
\end{equation}
is the noninteracting electron Green's function, $\xi_{\bm k}=\varepsilon_{\bm k} - E^0_\mathrm{F}$ is the noninteracting energy measured from the bare Fermi energy $E^0_\mathrm{F}$, $\eta$ denotes an infinitesimal positive number, $n_\mathrm{F}(x)$ is the Fermi-Dirac distribution function, and $W(\bm q,\nu)$ is the dynamically screened Coulomb interaction given in the RPA [Fig. 1(b)] by
\begin{equation}
W(\bm q,\nu)=\frac{v_c(\bm q)}{\varepsilon(\bm q,\nu)}
\end{equation}
where $v_c (q)=2\pi e^2/q$ is the 2D Coulomb interaction corresponding to the long-range $e^2/r$ interaction, and
\begin{equation}
\varepsilon(\bm q,\omega)=1-v_c(\bm q)\Pi_0(\bm q,\omega)
\label{eq:dielectric_function}
\end{equation}
is the two-dimensional dielectric function obtained within the RPA in which one sums up an infinite series of bubble diagrams [Fig.~\ref{fig:GW_diagram} (b)]. $\Pi_0(\bm q,\omega)$ is the noninteracting polarizability defined by the bare bubble and given by 
\begin{equation}
    \Pi_0(\bm q,\omega) = \int \frac{d^2k}{(2\pi)^2} \frac{n_\mathrm{F}(\xi_{\bm k})-n_\mathrm{F}(\xi_{\bm k+\bm q} )}{\omega+\varepsilon_{\bm k}-\varepsilon_{\bm k+\bm q}+i\eta}.
\end{equation}
The analytical expression for the polarizability of two dimensional isotropic electron gas is well-known \cite{Stern1967}, given by 
\begin{align}
    \Pi_0(\bm q,\omega)=&-\frac{m}{\pi} + \frac{m^2}{\pi q^2}
    \left[
    \sqrt{\left(    \omega+\frac{q^2}{2m}   \right)^2-\frac{2E^0_\mathrm{F} q^2}{2m}}\right.\nonumber \\
    &-
    \left.\sqrt{\left(    \omega-\frac{q^2}{2m}   \right)^2-\frac{2E^0_\mathrm{F} q^2}{2m}}\right],
    \label{eq:iso_polar}
\end{align}
where $q=\sqrt{q_x^2+q_y^2}$ and $E^0_\mathrm{F}$ is the bare Fermi energy. The polarizability for an anisotropic electron gas with unequal masses in an elliptic Fermi surface can be exactly obtained from Eq.~(\ref{eq:iso_polar}) by rescaling $m\rightarrow m_\mathrm{DOS}$, $q_x\rightarrow \sqrt{\frac{m_\mathrm{DOS}}{m_\mathrm{H}}}q_x$, and $q_y\rightarrow  \sqrt{\frac{m_\mathrm{DOS}}{m_\mathrm{L}}}q_y$. By rewriting the screened Coulomb interaction as
\begin{equation}
    W(\bm q,\nu)=v_c(\bm q)+v_c(\bm q)\left[\frac{1}{\varepsilon(\bm q,\nu)}-1\right],
\end{equation}
and putting it into Eq.~(\ref{eq:GW_selfenergy}), we can decompose the self-energy into $\Sigma({\bm k},\omega)=\Sigma^{\mathrm{ex}}({\bm k})+\Sigma^{\mathrm{corr}}({\bm k},\omega)$ where
\begin{align}
\Sigma^{\mathrm{ex}} ({\bm k})\!=&-\!\int\!\frac{d^2 q}{(2\pi)^2} n_\mathrm{F}(\xi_{\bm k+\bm q}) v_c(\bm q)
\label{eq:self_energy_ex}
\end{align}
is the frequency-independent exchange self-energy of the Hartree-Fock approximation, and 
\begin{equation}
    \Sigma^{\mathrm{corr}}({\bm k},\omega)\!=-\!\int\!\frac{d^2 q}{(2\pi)^2}
    \!\int\!\frac{d \nu}{2\pi i}
    \widetilde{W}(\bm q,\nu) 
    G_0(\bm k + \bm q, \nu+\omega)
    \label{eq:self_energy_corr}
\end{equation}
is the so-called correlation part that contains all the dynamical contributions beyond the static Hartree-Fock approximation of Eq. (9).
\begin{figure}[tb]
    \centering
    \includegraphics[width=0.7\linewidth]{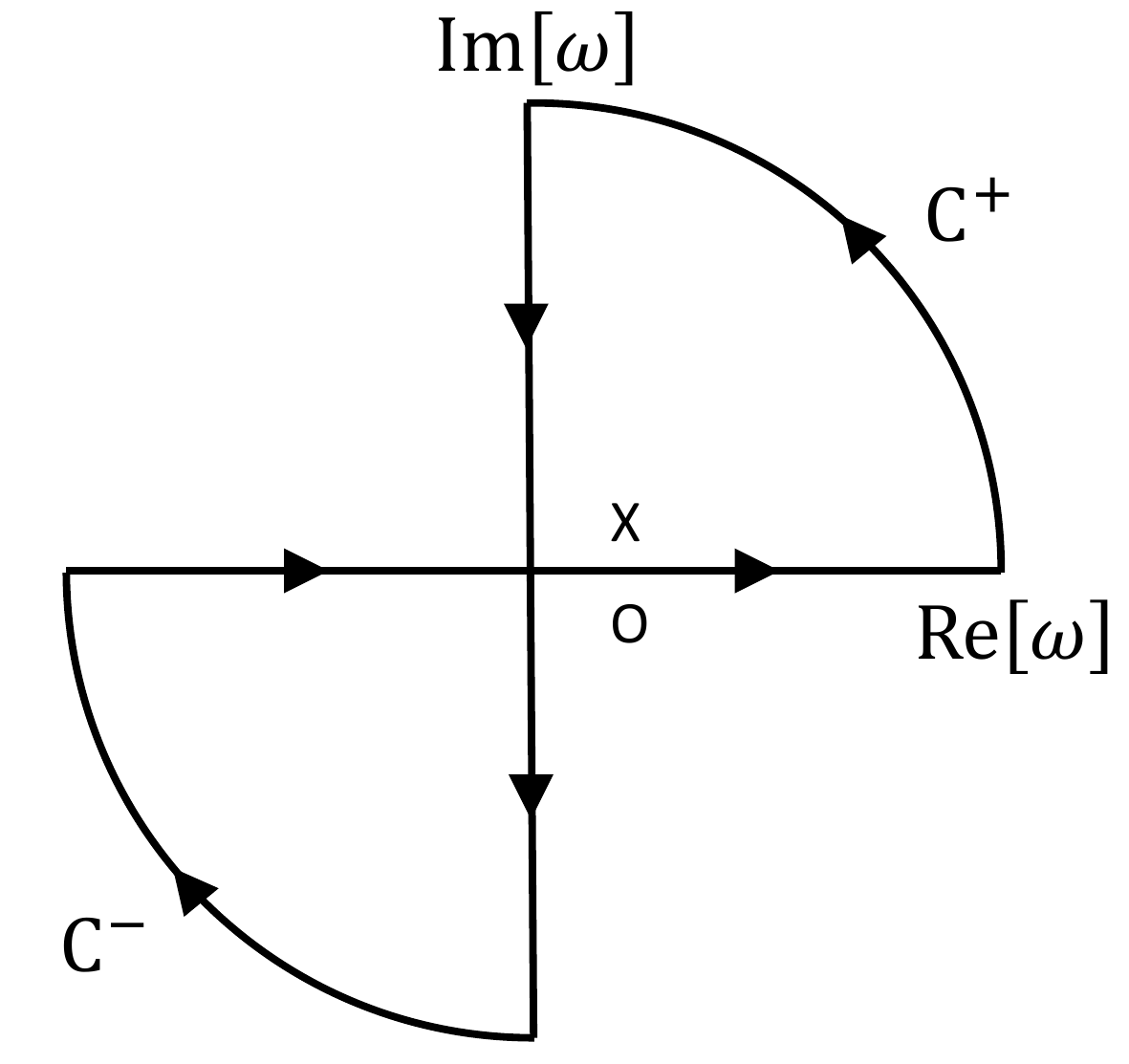}
    \caption{The integration contour is chosen to enclose the poles of $G_0(\bm k + \bm q, \nu+\omega)$, but not those of $W(\bm q,\nu)$. The cross and circle symbols represent the possible position of the pole of $G_0(q, \omega)$ and $W(\bm q,\nu)$, respectively.}
    \label{fig:contour}
\end{figure}
Here $\widetilde{W}(\bm q,\nu)$ is the screened Coulomb interaction minus the bare Coulomb interaction:
\begin{equation}
\widetilde{W}(\bm q,\nu)=W(\bm q,\nu)-v_c(\bm q)=v_c(\bm
q)\left[\frac{1}{\varepsilon(\bm q,\nu)}-1\right].
\label{eq:sc_bare}
\end{equation}
In obtaining $\Sigma^{\mathrm{ex}}$, we use $n_\mathrm{F}(\xi_{\bm k+\bm q})=\int \frac{d \nu}{2\pi i} G_0(\bm k + \bm q, \nu+\omega)$. 
Evaluating Eq.~(\ref{eq:self_energy_corr}) involves integration along the real frequency axis. Unfortunately, the integrand along the real axis has a complex singular structure arising from the poles and branch cuts of $\varepsilon(q,\omega)$, making a direct evaluation difficult. One of the ways to avoid this difficulty is to deform the integration path from the real axis to a contour in the complex plane. The contour path is chosen such that the path encloses only the poles of $G_0(\bm k + \bm q, \nu+\omega)$ but not those of $W(\bm q,\nu)$ as shown in Fig.~\ref{fig:contour}. The integration along the contour can be decomposed into four terms:
\begin{align}
    \oint \frac{d\nu}{2\pi i} &\widetilde{W}(\bm q,\nu) 
    G_0(\bm k + \bm q, \nu+\omega) \nonumber \\
    &=\int_{C^+}d\nu
    +\int_{C^-}d\nu
    +\int^{\infty}_{-\infty}d\nu
    -\int^{i\infty}_{-i\infty}d\nu.
    \label{eq:path}
\end{align}
The first two terms on the right-hand side cancel each other. Then the integration along the real frequency axis is expressed as the sum of the integration along the imaginary frequency axis and the contour integration, which can be evaluated by using the residue theorem. Since all the poles of $W(\bm q,\nu)$ are outside the contour path, we need to evaluate the residue only at the poles of $G_0(\bm k + \bm q, \nu+\omega)$:
\begin{align}
    \oint &\frac{d\nu}{2\pi i}\widetilde{W}(\bm q,\nu)G_0(\bm k + \bm q, \nu+\omega)=\nonumber \\
    &-[1-n_\mathrm{F}(\xi_{\bm {k+q}})]\Theta(\omega-\xi_{\bm {k+q}})\widetilde{W}(\bm q,\xi_{\bm {k+q}}-\omega-i\eta) \nonumber\\
    &+n_\mathrm{F}(\xi_{\bm {k+q}})\Theta(\xi_{\bm {k+q}}-\omega)\widetilde{W}(\bm q,\xi_{\bm {k+q}}+\omega+i\eta),
    \label{eq:residue}
\end{align}
where $\Theta(x)$ is the step function [$\Theta(x)=1$ for $x>0$ and $0$ otherwise]. 
Using Eqs.~(\ref{eq:self_energy_corr})-(\ref{eq:residue}), we obtain $\Sigma^{\mathrm{corr}}({\bm k},\omega)=\Sigma^{\mathrm{line}}({\bm k},\omega)+\Sigma^{\mathrm{res}}({\bm k},\omega)$ where
\begin{align}
\Sigma^{\mathrm{line}} ({\bm k},\omega)\!=&-\!\int\!\frac{d^2 q}{(2\pi)^2}\int_{-\infty}^{\infty}\!\frac{dv}{2\pi} 
\frac{v_c(\bm q)}{\omega+iv-\xi_{\bm k+\bm q}} \nonumber \\ 
&\times\left[\frac{1}{\varepsilon(\bm q,i v)}-1\right]
\label{eq:self_energy_line}
\end{align}
is from the integration along the imaginary axis and
\begin{align}
\Sigma^{\mathrm{res}} ({\bm k},\omega)\!=&\!\int\!\frac{d^2 q}{(2\pi)^2} \left [\Theta(\omega-\xi_{\bm k+\bm q}) - \Theta(-\xi_{\bm k+\bm q}) \right ] \nonumber \\ 
&\times v_c(\bm q)\left[\frac{1}{\varepsilon(\bm q,\xi_{\bm k+\bm q}-\omega)}-1\right]
\label{eq:self_energy_res}
\end{align}
from the contour integration using the residue theorem. Note that since $\varepsilon^*(\bm q,i\nu)=\varepsilon(\bm q,-i \nu)=\varepsilon(\bm q,i \nu)$, $\Sigma^\mathrm{line}$ is always real. $\Sigma^\mathrm{ex}$ is also real as can be easily seen from Eq.~(\ref{eq:self_energy_ex}). Thus, the imaginary part of the self-energy comes entirely from $\Sigma^\mathrm{res}$, i.e., $\mathrm{Im}\Sigma=\mathrm{Im}\Sigma^{\mathrm{res}}$. The exchange self-energy of the Hartree-Fock theory [Eq.~\ref{eq:self_energy_ex}] is pure real, and does not contain any decoherence mechanism arising from electron-electron interactions. We emphasize that the self-energy described in this sub-section is exact in the high-density ($r_s<1$) limit since the infinite series of the bubble diagrams of the RPA [Fig.~\ref{fig:GW_diagram}(b)] are the most divergent diagrams arising from Coulomb interaction in each order at high electron density.

\subsection{Plasmon-pole approximation}
Although we develop the full anisotropic RPA theory for the 2D self-energy in this paper, we start with a simple model, the plasmon-pole approximation, which is easy to use and is quantitatively often an adequate approximation, accurately reproducing the RPA self-energy results.

The spirit of the PPA is to simplify the calculation of the correlation part of the self-energy [Eq.~(\ref{eq:self_energy_corr})] by replacing the full dynamical dielectric function with the effective dielectric function having a single plasmon mode, given by \cite{Hwang2018, Vinter1975, Vinter1976, Lundqvist1967, Lundqvist1967a}
\begin{equation}
    \frac{1}{\varepsilon(\bm q,\omega)} -1  = 
    \frac{\omega^2_p(\bm q)}{\omega^2 - \omega_{\bm q}^2 - i\eta}
    \label{eq:PPA_dielectric}
\end{equation}
that satisfies the $f$-sum rule
\begin{equation}
    \int^\infty_0 d\omega \frac{1}{\omega} \mathrm{Im}\left[\frac{1}{\varepsilon(\bm q,\omega)} -1  \right]=-\frac{\pi}{2}\frac{\omega^2_p(\bm q)}{\omega_{\bm q}^2}
\end{equation}
where $\omega_p(\bm q)$ is the plasma frequency in the long-wavelength limit and $\omega_{\bm q}$ is the effective PPA plasma frequency. Using the Kramers-Kronig relation
\begin{equation}
    \int^\infty_0 d\omega \frac{1}{\omega} \mathrm{Im}\left[\frac{1}{\varepsilon(\bm q,\omega)} -1  \right]=\frac{\pi}{2}\left[\frac{\omega_p^2(\bm q)}{1/\varepsilon(\bm q,0)-1} \right],
\end{equation}
we find that
\begin{equation}
    \omega_{\bm q}^2=-\frac{\omega^2_p(\bm q)}{1/\varepsilon(\bm q,0)-1}.
\end{equation}
By putting Eq.~(\ref{eq:PPA_dielectric}) into  Eq.~(\ref{eq:self_energy_corr}), and performing the frequency integration, which can be performed analytically in contrast to the RPA case, the correlation self-energy within the PPA becomes
\begin{align}
\Sigma^\mathrm{corr}({\bm k},\omega)\!=&i\!\int\!\frac{d^2 q}{(2\pi)^2}
\frac{v_c(\bm q)\omega_p^2}{2\omega_{\bm q}}
\left[
\frac{\Theta(-\xi_{\bm k+ \bm q})}{\omega+\omega_{\bm q}-\xi_{\bm k+\bm q}-i\eta}\right.
\nonumber \\ 
&+\left.\frac{\Theta( \xi_{\bm k+ \bm q})}{\omega-\omega_{\bm q}-\xi_{\bm k+\bm q}+i\eta}
\right].
\end{align}
The long wavelength plasma frequency $\omega_\mathrm{p}(\bm q)$ can be obtained as follows. We first expand the polarizability to the first order of $q$, 
\begin{equation}
    \Pi_0(\bm q,\omega)\approx \frac{m_\mathrm{DOS}}{\pi}\frac{2\widetilde{q}^2}{\widetilde{\omega}}\left( \sqrt{\frac{m_\mathrm{L}}{m_\mathrm{H}}}\cos^2\theta + \sqrt{\frac{m_\mathrm{H}}{m_\mathrm{L}}}\sin^2\theta  \right),
    \label{eq:polarizability}
\end{equation}
where $\widetilde{q}=q/k^0_\mathrm{F}$, $\widetilde{\omega}=\omega/E^0_\mathrm{F}$ and $k^0_\mathrm{F}=\sqrt{2m_\mathrm{DOS}E^0_\mathrm{F}}$. Here we use the polar coordinate: $q_x=q \cos{\theta}$, $q_y=q \sin{\theta}$.
The plasmon dispersion is given by the zeros of the dynamical dielectric function.
Thus, by putting Eq.~(\ref{eq:polarizability}) into Eq.~(\ref{eq:dielectric_function}) and solving the equation $\varepsilon[\bm q,\omega_p(\bm q)]=0$, we obtain $\omega_p(\bm q)$ to be
\begin{equation}
    \widetilde{\omega}_p(\bm q)=2^{3/4}\sqrt{\widetilde{q} r_s\left( \sqrt{\frac{m_\mathrm{L}}{m_\mathrm{H}}}\cos^2\theta + \sqrt{\frac{m_\mathrm{H}}{m_\mathrm{L}}}\sin^2\theta  \right)}.
\end{equation}

The PPA is extensively used since it has been proven to be an excellent approximation successfully describing many-body effects for various systems \cite{Vinter1975, Vinter1976,Sarma1996,Hwang2018}. In each of the following sections, we compare results obtained in the RPA with those in the PPA in order to see how well the PPA works for an electron gas with anisotropic effective mass. The advantage of PPA is its great ease of use, considerably simplifying the self-energy calculations.  The disadvantage is that it is an uncontrolled approximation to the RPA, so its accuracy in a particular situation is not known a priori.

\begin{figure*}[!htb]
    \centering
    \includegraphics[width=1\linewidth]{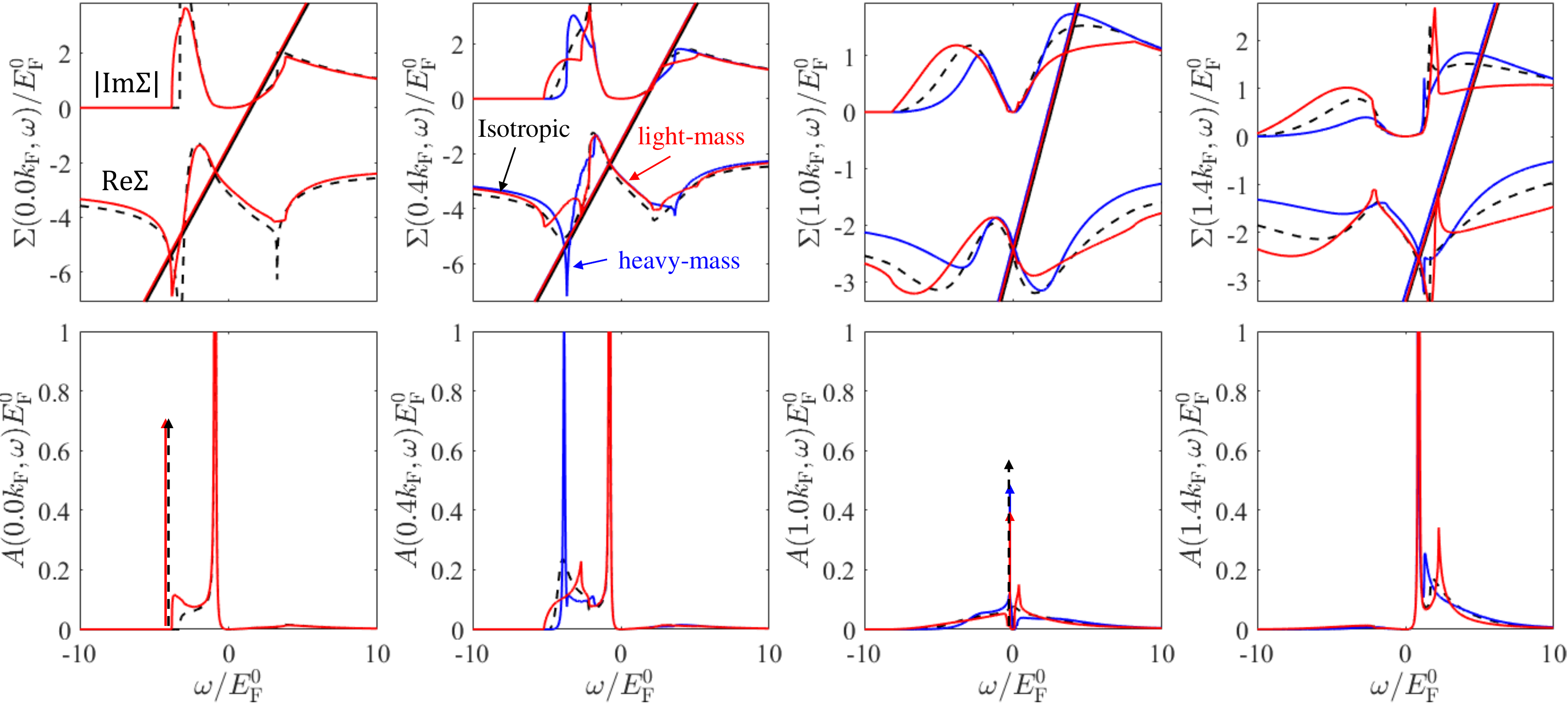}
    \caption{The upper four figures show the numerically calculated real and imaginary part of self-energies along the heavy-mass (blue) and light-mass (red) directions for fixed momenta $k/k_\mathrm{F}=0.0$, $0.4$, $1.0$, and $1.4$. We plot $|\mathrm{Im}\Sigma|$ instead of $\mathrm{Im}\Sigma$ for clarity of visual representation. $k_\mathrm{F}$ is the magnitude of the renormalized Fermi wavevector along the corresponding direction, i.e., $k_\mathrm{F}=k_{\mathrm{F}_x}$($k_{\mathrm{F}_y}$) for the heavy-mass (light-mass) direction, which can be obtained by solving Eq.~(\ref{eq:interacting_mu_final}). The black dashed line represents the isotropic self-energy calculated using the corresponding density-of-states mass $m_{\mathrm{DOS}}$. The straight lines are given by $\omega - \varepsilon_{\bm k} + E_\mathrm{F}$, whose intersection with $\mathrm{Re}\Sigma$ provides the solutions of the Dyson's equation. The lower four figures show the spectral functions extracted from the corresponding self-energies. The vertical arrows represent $\delta$-function peaks and their heights do not reflect the spectral weights. Here we set $m_\mathrm{H}/m_\mathrm{L}=10$ and $r_s=2.0$ (see Appendix \ref{sec:appendixB} for results for different values of $r_s$ and mass ratios).}
    \label{fig:self_energy_smallk}
\end{figure*}

\section{Quasiparticle Self-Energy} \label{sec:self_energy}
The retarded interacting Green's function for an anisotropic electron gas is written as
\begin{equation}
    G(\bm k,\omega) = \frac{1}{\omega - \varepsilon_{\bm k} + E_\mathrm{F}- \Sigma(\bm k,\omega)}.
\end{equation}
The quasiparticle energy $E(\bm k)$ is determined by the pole of the retarded interacting Green's function, leading to the Dyson's equation
\begin{equation}
    E(\bm k) - \varepsilon_{\bm k} + E_\mathrm{F} = \mathrm{Re}\Sigma[\bm k,E(\bm k)].
    \label{eq:Dyson}
\end{equation}
We can obtain the renormalized Fermi energy $E_\mathrm{F}$ and Fermi wavevectors $\bm k_\mathrm{F}$ by setting $\bm k=\bm k_\mathrm{F}$ in Eq.~(\ref{eq:Dyson}), leading to a self-consistent equation given by 
\begin{equation}
    E_\mathrm{F}=\varepsilon_{\bm k_\mathrm{F}} + \mathrm{Re}\Sigma(\bm{k}_\mathrm{F},0).
    \label{eq:interacting_mu}
\end{equation}
For an isotropic circular Fermi surface, Eq.~(\ref{eq:interacting_mu}) is easily solved because the renormalized Fermi surface is the same as its noninteracting counterpart due to the Luttinger theorem stating that the volume enclosed by the Fermi surface is proportional to the electron density \cite{Luttinger1955}. For an anisotropic Fermi surface, however, there are many possible Fermi surface shapes of the same volume and thus the Luttinger theorem does not guarantee that the renormalized Fermi surface remains unaltered by interactions unlike the case for an isotropic Fermi surface. 
The exact shape of the distorted interacting Fermi surface can be obtained by finding a set of $\bm k_\mathrm{F}$ and $E_\mathrm{F}$ satisfying Eq.~(\ref{eq:interacting_mu}) and the Luttinger's theorem simultaneously. This is quite a formidable task. 
It has recently been shown, however, that the deviation of the interacting Fermi surface from an elliptical shape is negligibly small for a wide range of $r_s$ \cite{Ahn2020}, justifying to approximate the Fermi surface as an ellipse characterized by only the magnitude of the two Fermi wave vectors along the principal axes, i.e., $k_{F_x}$ and $k_{F_y}$. We use this elliptic approximation with renormalized effective mass for most of the results presented in this paper, emphasizing that this approximation is excellent as the shape deviation of the interacting Fermi surface from an ellipse is less than 1$\%$ for all the cases presented in this paper (We provide some results for the interacting Fermi surface topology in Appendix A). Within the elliptical approximation, Eq.~(\ref{eq:interacting_mu}) is reduced to the following two equations in terms of $E_\mathrm{F}$, $k_{\mathrm{F}x}$ and $k_{\mathrm{F}y}$:
\begin{equation}
\begin{aligned}
    E_\mathrm{F}&=\varepsilon_{k_{\mathrm{F}x}} + \mathrm{Re}\Sigma({k_{\mathrm{F}x}},0),\\
    E_\mathrm{F}&=\varepsilon_{k_{\mathrm{F}y}} + \mathrm{Re}\Sigma({k_{\mathrm{F}y}},0).
    \label{eq:interacting_mu_final}
\end{aligned}
\end{equation}

\begin{figure*}[tb]
    \centering
    \includegraphics[width=1\linewidth]{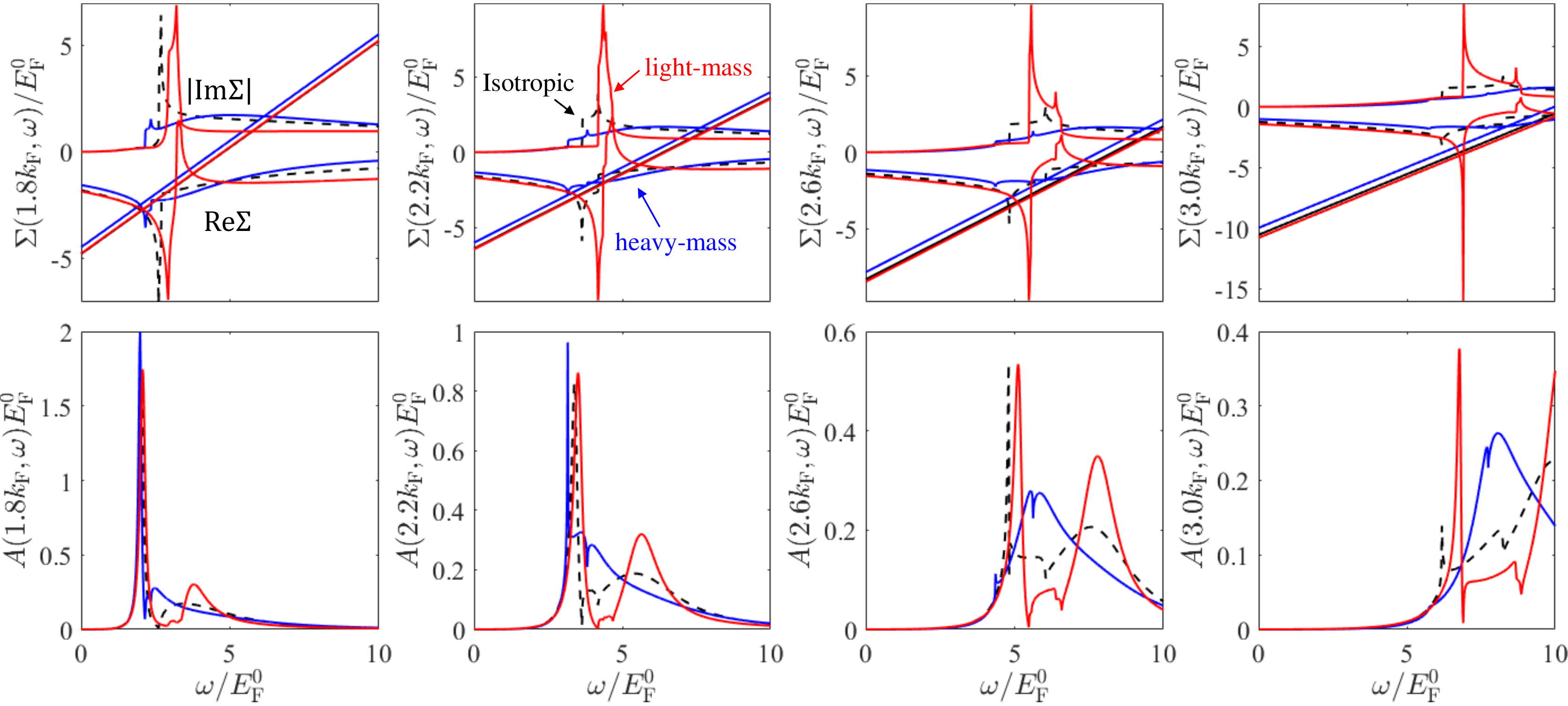}
    \caption{Calculated self-energies and the corresponding spectral functions for larger wavevectors $k/k_\mathrm{F}=1.8$, $2.2$, $2.6$, and $3.0$. The notation is the same as in Fig.~\ref{fig:self_energy_smallk}}
    \label{fig:self_energy_largek}
\end{figure*}

The upper four figures in Fig.~\ref{fig:self_energy_smallk} show calculated self-energies as a function of $\omega$ for several fixed wavevectors along the heavy-mass (blue line) and the light-mass (red line) directions. The black dashed line represents the isotropic self-energy calculated using the corresponding density-of-states mass. The intersection between the straight line and $\mathrm{Re}\Sigma$ represents the solution of the Dyson's equation [Eq.(\ref{eq:Dyson})], and each solution gives rise to a peak in the single-particle spectral function defined as 
\begin{equation}
    A(\bm k,\omega)=\frac{1}{\pi} \frac{|\mathrm{Im}\Sigma(\bm k,\omega)|}{[\omega+E_\mathrm{F}-\varepsilon_{\bm k}-\mathrm{Re}\Sigma(\bm k,\omega)]^2+[\mathrm{Im}\Sigma(\bm k,\omega)]^2},
\end{equation}
which gives the probability of finding an electron with momentum $\bm k$ and energy $\omega$. The sharpness of the spectral function defines the sharpness of the quasiparticle, but the strict quasiparticle with zero damping (i.e., a $\delta$-function spectral function) exists only on the Fermi surface.

The isotropic result (black dashed lines) for $k=0$ shows the typical behavior of the self-energy and spectral function: There are three solutions of Dyson's equations. The solution closest to $\omega=0$ corresponds to the usual Landau quasiparticle excitation with an energy slightly shifted (``renormalized'') from its noninteracting value. Note that the quasiparticle solution shows up as a sharp peak in the spectral function because its damping defined by $\mathrm{Im}\Sigma$ is small. One of the two remaining solutions is heavily suppressed due to the large $\mathrm{Im}\Sigma$ whereas the other one is undamped, giving rise to a strong peak in the spectral function. This peak is called a plasmaron and interpreted as a composite particle formed by the coupling between electrons and plasmons \cite{Lundqvist1967, Lundqvist1967a, Lundqvist1970}. 
\begin{figure*}[tb]
    \centering
    \includegraphics[width=1\linewidth]{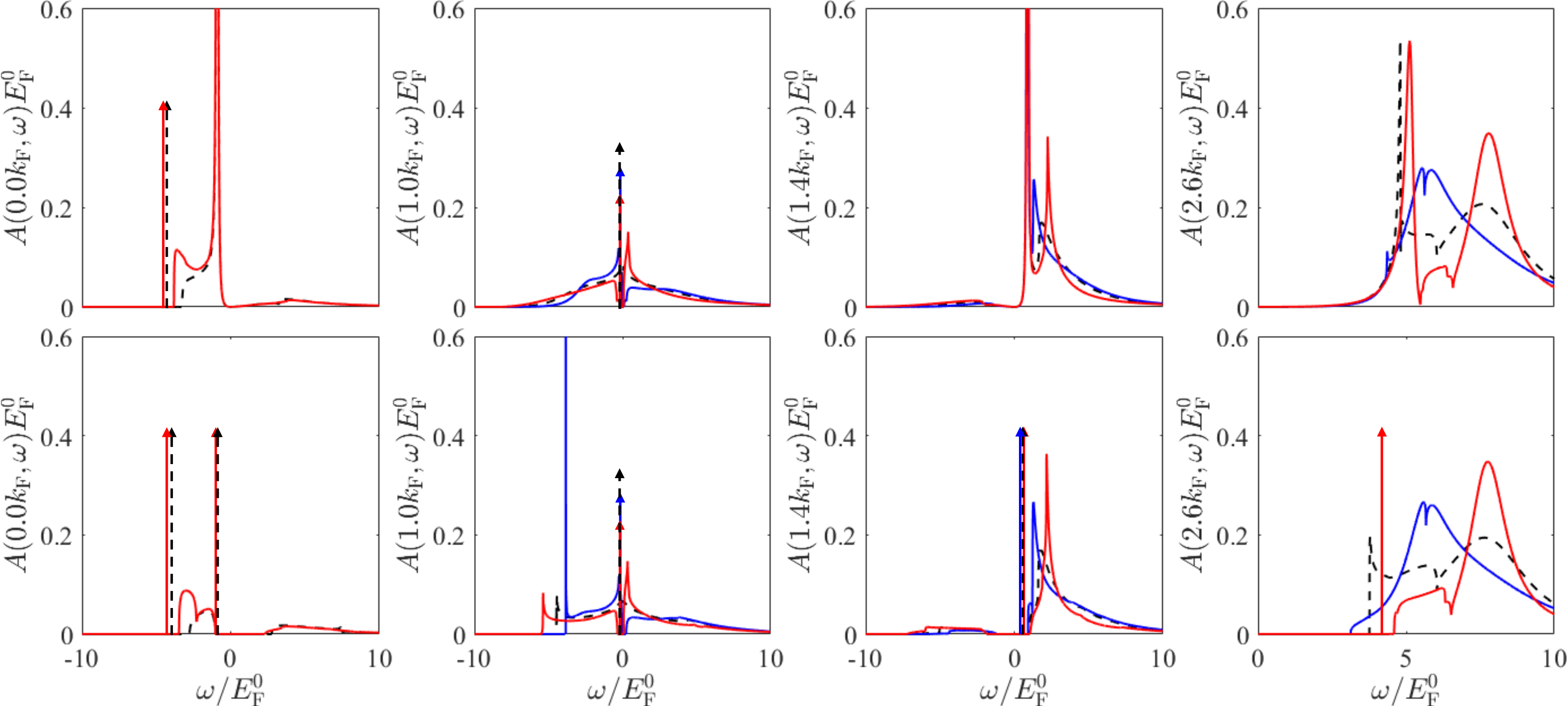}
    \caption{Spectral functions calculated within the RPA (upper four figures) and PPA (lower four figures). The notation is the same as in Fig.~\ref{fig:self_energy_smallk} }
    \label{fig:self_energy_PPA}
\end{figure*}

For $k=0$, the self-energy for the anistropic system (red line) and the one calculated using the isotropic density-of-states mass (black dahsed line) are in good agreement, resulting in an almost identical structure of the spectral function.
This indicates that the isotropic approximation using $m_\mathrm{DOS}$ works well for $k=0$.
For $k=0.4k_\mathrm{F}$, the anisotropic self-energies along the heavy- (blue) and light-mass (red) directions are in good agreement around the the Dyson's equation solution giving rise to the quasiparticle peak near $\omega=0$, and well approximated by the isotropic self-energy. Around the plasmaron solution, however, $\mathrm{Re}\Sigma$ along the light-mass direction shows a distinct behavior, exhibiting a double-peak structure which is absent in the other two curves. Due to this structure, the plasmaron solution is shifted to a lower frequency where $\mathrm{Im}\Sigma$ is large, and thus the corresponding plasmaron peak in the spectral function is much more suppressed compared to the one along the heavy-mass direction. 
At the Fermi wavevector ($k=k_\mathrm{F}$), the anisotropic spectral function is asymmetric around $\omega=0$ in contrast to the isotropic one which is symmetric. But both the isotropic and the anisotropic spectral functions show the well-known delta-function quasiparticle peak at $\omega=0$ because of the vanishing $\mathrm{Im}\Sigma$ behaving as $\mathrm{Im}\Sigma(k_\mathrm{F},\omega)\approx\omega^2\ln(\omega)$ as $\omega\rightarrow0$, which is a universal property of 2D Fermi liquids \cite{PhysRevB.53.9964}.
For $k=1.4k_\mathrm{F}$, the structure of $\mathrm{Re}\Sigma$ along the light-mass direction is similar to that of $\mathrm{Re}\Sigma$ along the heavy-mass direction near the quasiparticle solution, giving rise to an almost identical quasiparticle peak in the spectral function. It should be noted, however, that $\mathrm{Re}\Sigma$ along the light-mass direction exhibits a singularity much stronger than those in the other two curves, and it shifts the plasmaron solution to a higher energy compared to the one along the heavy-mass direction.   
The isotropic results obtained using the density-of-states mass duplicate the anisotropic results quite well around the quasiparticle solution. However, the isotropic self-energy seriously deviates from the anisotropic self-energy around the plasmaron solution, and thus fails to capture the anisotropic feature of the plasmaron peak.

Figure~\ref{fig:self_energy_largek} shows calculated self-energies for larger wavevectors $k/k_\mathrm{F}=1.8$, $2.2$, $2.6$, and $3.0$. Note that the singular structure in $\mathrm{Re}\Sigma$ along the light-mass direction becomes stronger with increasing wavevector $k$. When the singularity is strong enough, the self-energy along the light-mass direction behaves quite differently from that along the heavy-mass direction near the quasiparticle solution, leading to a separation of the quasiparticle peaks along the two different directions in the spectral function.
Considering that the quasiparticle concept breaks down as one moves away from the from the Fermi surface (i.e., at large wavevectors), it is worthwhile to look closely at the evolution and disappearance of the quasiparticle peak with increasing wavevectors $k$.
Comparing among the results for $k/k_\mathrm{F}=2.2$,$2.6$ and $3.0$, one can immediately notice that the quasiparticle peak along the heavy-mass direction rapidly disappears as we increase $k$ from $2.2k_\mathrm{F}$ to $3.0k_\mathrm{F}$ while the quasiparticle peak along the light-mass direction remains sharp up to $k=3.0k_\mathrm{F}$. Such an anisotropic suppression of the quasiparticle peak can be understood by looking at the corresponding self-energy:
The real part of the self-energies for $k/k_\mathrm{F}=2.6$ and $3.0$ show that along the heavy-mass direction there is only one solution to the Dyson's equation that corresponds to the incoherent dispersive peak. Along the light-mass direction, however, the strong singularity in $\mathrm{Re}\Sigma$ allows three Dyson equation solutions, only one of which near $\omega=0$ gives rise to a well-defined quasiparticle peak in the spectral function. Obviously, any isotropic approximation misses these key features of the anisotropic quasiparticles away from the Fermi surface.

Figure~\ref{fig:self_energy_PPA} shows the spectral functions calculated within the RPA (top four figures) and the PPA (bottom four figures). Note that the RPA and PPA self-energies are in an impressive qualitative agreement despite the simplicity of the PPA, capturing all the anisotropic features observed in the RPA results discussed above. However there are a few discrepancies that should be noted: For $k=0$, $1.4k_\mathrm{F}$ and $2.6k_\mathrm{F}$, the PPA gives an ideal delta function quasiparticle peak in contrast to the RPA results where the quasiparticle peaks are weakly damped. In addition the PPA spectral function for the Fermi wavevector ($k=k_\mathrm{F}$) exhibits a spurious peak located below the Fermi energy, which is known to occur because the PPA overestimates the plasmaron feature \cite{Jalabert1989}. These undesirable features are already present in the isotropic PPA theory, and are intrinsic to the plasmon-pole approximation.

\section{Scattering Rate} \label{sec:scattering_rate}
In this section, we investigate the effect of mass anisotropy on the interaction-induced inelastic scattering rate. The scattering rate can be calculated using the imaginary part of the self-energy via the relation $\Gamma=2\mathrm{Im}\Sigma(\bm k,\xi_{\bm k})$, and we can obtain $\mathrm{Im}\Sigma(\bm k,\xi_{\bm k})$ using Eq.~(\ref{eq:self_energy_res}):
\begin{align}
    \mathrm{Im}\Sigma ({\bm k},\xi_{\bm k})\!=&\!\int\!\frac{d^2 q}{(2\pi)^2} \left [\Theta(\xi_{\bm k}-\xi_{\bm k+\bm q}) - \Theta(-\xi_{\bm k+\bm q}) \right ] \nonumber \\ 
    &\times v_c(\bm q)\mathrm{Im}\left[\frac{1}{\varepsilon(\bm q,\xi_{\bm k+\bm q}-\xi_{\bm k})}\right].
    \label{eq:scattering_rate}
\end{align}
Here, following the standard convention, we use the on-shell approximation to define the scattering rate where $\omega$ in the self-energy is substituted by the on-shell energy $\xi_{\bm k}$.

\begin{figure*}[!htb]
    \centering
    \includegraphics[width=0.9\linewidth]{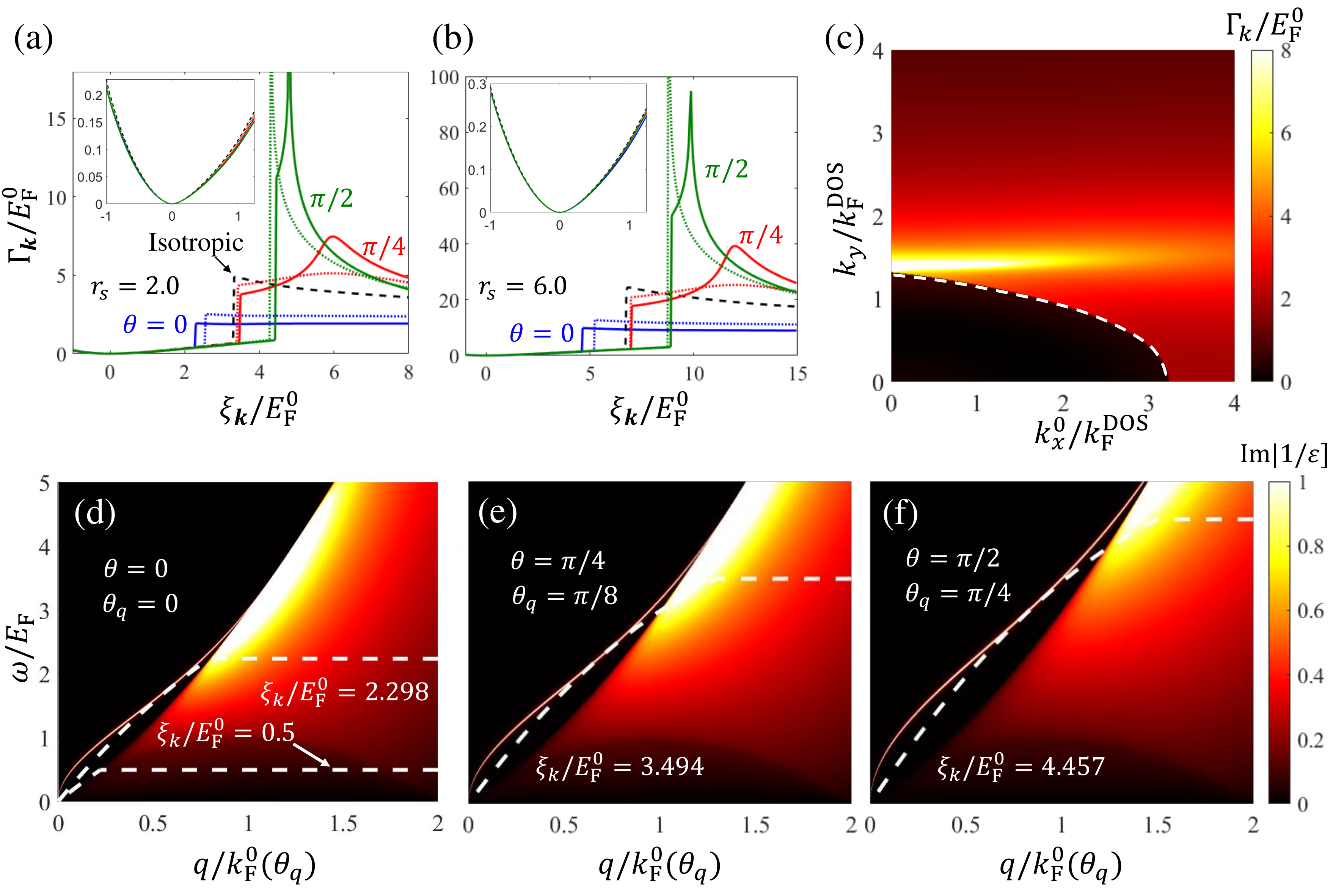}
    \caption{(a) and (b) Calculated scattering rates for the mass ratio $m_\mathrm{H}/m_\mathrm{L}=5$ (dotted) and $10$ (solid) along three different directions, which are represented by the angle $\theta$ from the heavy-mass axis. The black dashed line represents the isotropic scattering rate with the density-of-states mass. The insets show the scattering rate near the Fermi surface. (c) Two-dimensional plot of the scattering rate for $m_\mathrm{H}/m_\mathrm{L}=10$. The white dashed line represents the momenta of the injected electron at which an abrupt jump in the scattering rate associated with plasmon emissions occur. Here $k_\mathrm{F}^\mathrm{DOS}=\sqrt{2m_\mathrm{DOS}E^0_F}$ (d)-(f) the energy loss function for $r_s=2.0$ and $m_\mathrm{H}/m_\mathrm{L}=10$ plotted along the direction of the scattering transfer momentum $\bm q$, whose direction is indicated in the figure by the angle $\theta_q=\tan^{-1}(q_y/q_x)$. $k^0_{\mathrm F}(\theta_q)$ is the magnitude of the noninteracting Fermi wavevector along the direction at the angle of $\theta_q$, and the white dashed line represents the boundary of the injected-electron energy-loss (IEEL) continuum for an electron injected with energies (d) $\xi_k/E^0_\mathrm{F}=0.5$ and $2.298$, (e) $\xi_k/E^0_\mathrm{F}=3.494$, and (f) $\xi_k/E^0_\mathrm{F}=4.457$, } 
    \label{fig:scattering_rate}
\end{figure*}

Figure~\ref{fig:scattering_rate} (a) shows the calculated scattering rates along the directions at the angles of $\theta\coloneqq\tan^{-1}(k_y/k_x)=0$ (blue), $\pi/4$ (red), and $\pi/2$ (green) from the heavy-mass axis as a function of the on-shell energy $\xi_{\bm k}$. Near the Fermi surface (i.e., $\xi_{\bm k}\rightarrow0$), the scattering rates along the three different directions vanish at an almost same rate showing the well-known behavior of $\sim\xi_{\bm k}^2\ln{\left|\xi_{\bm k}\right|}$, and are in good agreement with the isotropic scattering rate calculated using the density-of-states mass. Away from the Fermi surface, however, the scattering rate becomes strongly anisotropic, exhibiting an abrupt jump at different energies depending on the direction $\theta$.
Note that the scattering rate along the heavy-mass direction ($\theta=0$) monotonically decreases after the threshold energy similar to the isotropic scattering rate. Along the other two directions, however, the scattering rate shows a different behavior: the scattering rate along the direction $\theta=\pi/4$ continues to increase after the threshold energy, reaches a maximum and then monotonically decreases. Along the light-mass direction ($\theta=\pi/2$) the scattering rate has a sharp peak and a kink structure (for $m_\mathrm{H}/m_\mathrm{L}=10$) around the threshold energy. Both of these behaviors for $\theta=\pi/4$ and $\theta=\pi/2$ are absent in the isotropic scattering rate, and thus cannot be qualitatively predicted by the isotropic approximation. To investigate how changing $r_s$ affects the scattering rate, we plot the calculated scattering rate for a larger $r_s=6.0$ in Fig~\ref{fig:scattering_rate} (b). Note that for a given mass ratio, the main quantitative change in the scattering rate with increasing $r_s$ is in its scale, but the gross qualitative behavior remains unaltered. The anisotropy of the scattering rate is highlighted in Fig.~\ref{fig:scattering_rate} (c), where we show a two dimensional color plot of the scattering rate for $m_\mathrm{H}/m_\mathrm{L}=10$ and $r_s=2.0$ as a function of $k_x$ and $k_y$.

In the following we provide an analysis of the decay process involved in the anisotropic scattering rate. Figures~\ref{fig:scattering_rate} (d) shows the loss function $\left|\mathrm{Im}[1/\varepsilon(q,\omega)]\right|$ along the heavy-mass direction, along with the boundary of the IEEL continua, which is the phase space allowed for the scattering processes and is mathematically equivalent to the set of ($\bm q$, $\omega=\xi_{\bm k}-\xi_{\bm k+\bm q}$) satisfying $\Theta(\xi_{\bm k}-\xi_{\bm k+\bm q}) - \Theta(-\xi_{\bm k+\bm q})\neq0$ so that $\mathrm{Im}[\Sigma(\bm k, \xi_{\bm k})]$ is nonzero. $\bm k$ is interpreted as the momentum of the injected electron, and $\bm q$ is the momentum transfer of the scattering. The loss function $\mathrm{Im}[1/\varepsilon(q,\omega)]$ in Eq.~(\ref{eq:scattering_rate}) describes energy dissipation via electron-hole pair excitations and its poles represent the dissipation via plasmon emission. Thus the intersection of the phase space with the electron-hole continua and with the plasmon dispersion indicates that the quasiparticle decay occurs via the emission of the electron-hole pairs and plasmons, respectively.
For an electron injected with low energy, the IEEL continuum covers only the electron-hole continuum [see the IEEL continuum for $\xi_{\bm k}/E_\mathrm{F}=0.5$ in Fig~\ref{fig:scattering_rate} (d)], and thus at small momenta the only available quasiparticle decay channel is through the emission of electron-hole pairs. With increasing $\xi_{\bm k}$, the IEEL continuum grows and covers the plasmon dispersion when $\xi_{\bm k}$ reaches the threshold energy [see the IEEL continuum for $\xi_{\bm k}/E_\mathrm{F}=2.298$ in Fig~\ref{fig:scattering_rate} (d)], turning on an additional decay channel via plasmon emission. This leads to an abrupt upward increase in the scattering rates seen in Figs.~\ref{fig:scattering_rate} (a) and (b). 

Figures~\ref{fig:scattering_rate} (e) and 6(f) show the loss functions and the IEEL continua corresponding to the threshold energy of plasmon emissions plotted along the two other remaining directions [(e) $\theta=\pi/4$ and (f) $\theta=\pi/2$]. Note that in our scaled unit the electron-hole and IEEL continua are identical along different directions, whereas the plasmon energy dispersions are not. This leads to a direction-dependent threshold energy for the quasiparticle decay into plasmons. This direction dependence is a direct many-body manifestation of mass anisotropy, which is absent in any isotropic approximation. Recent work has shown that the 2D plasmon dispersion and decay (from Landau damping into electron-hole pairs) manifest qualitatively novel and quantitatively important anisotropy effects with no analogs in the corresponding isotropic 2D system \cite{Ahn2020plasmons}. Our results in the current paper establish that the reverse effect is also present in the inelastic damping of energetic quasiparticles due to 2D plasmon emission, which has characteristic anisotropic features with no analogs in the corresponding isotropic system. Thus, damping and decay of both quasiparticles and collective modes manifest nontrivial anisotropic inelastic features which cannot be captured in any isotropic approximation.

\begin{figure}[!htb]
    \centering
    \includegraphics[width=0.9\linewidth]{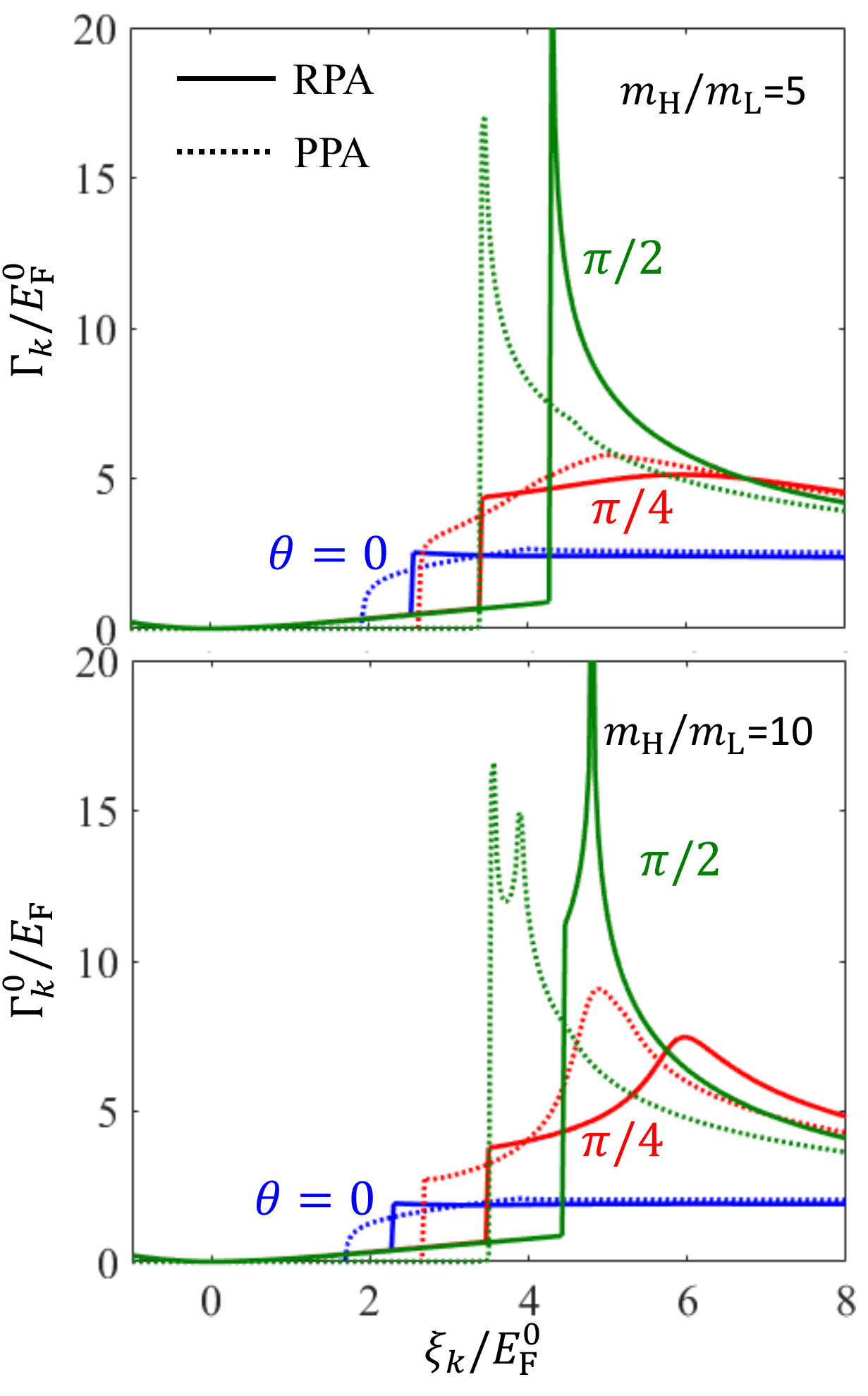}
    \caption{Scattering rates calculated within the RPA (solid) and PPA (dotted) for $m_\mathrm{H}/m_\mathrm{L}=5$ (upper) and $10$ (lower).   }
    \label{fig:scattering_rate_PPA}
\end{figure}

Figure~\ref{fig:scattering_rate_PPA} shows the scattering rates obtained within the PPA (dotted line) in comparison with those within the RPA (solid line). The PPA is a poor approximation for the inelastic scattering rate at low energies near $\xi_k=0$ where the contribution from electron-hole pair excitations is dominant because PPA emphasizes the plasmon modes. As in the RPA results, however, the PPA scattering rate abruptly increases at a certain threshold plasmon emission energy that varies depending on the direction $\theta$. Also note that the gross qualitative behaviors of the RPA and PPA scattering rates above the threshold energy are very similar. This agreement is because the PPA accurately takes into account the anisotropy of the plasmon dispersion through the effective dielectric function [Eq.~(\ref{eq:PPA_dielectric})] and thus provides an excellent approximation to the scattering rate when the decay process via plasmon emissions is dominant. PPA is, however, a poor approximation when the scattering arises from electron-hole excitations independent of any anisotropy because PPA involves approximating the full dynamical RPA screening by an effective plasmon pole thus undermining the electron-hole contributions.

\section{Effective Mass} \label{sec:effective_mass}
In this section, we present results for the renormalized effective masss in the anisotropic system. We first begin with the basic formalism for the many-body effective mass renormalization.
Assuming that the renormalized Fermi surface has an elliptical shape, we can write the rernormalized energy dispersion  
\begin{equation}
    E(\bm k)=E(0)+\frac{k_x^2}{2m^*_\mathrm{H}}+\frac{k_y^2}{2m^*_\mathrm{L}},
\label{eq:renorm_energy}
\end{equation}
where $m^*_\mathrm{L}$ and $m^*_\mathrm{H}$ denote the renormalized effective masses. Expanding the renormalized energy dispersion around the Fermi surface, we obtain
\begin{equation}
    E(\bm k)\approx E_\mathrm{F}+(k_x-k_{\mathrm{F}_x})\frac{k_{\mathrm{F}_x}}{m^*_\mathrm{H}} + (k_y-k_{F_y})\frac{k_{F_y}}{m^*_\mathrm{L}}.
\label{eq:renorm_energy_expanded}
\end{equation}
 Using Eq.~(\ref{eq:renorm_energy_expanded}), we can find an expression for the renormalized effective mass given by
\begin{equation}
\begin{aligned}
    m^*_\mathrm{H}&=k_{\mathrm{F}_x}\left.\left(\frac{\partial E(\bm k)}{\partial k_x}\right)^{-1}
    \right\rvert_{\bm k = \bm k_\mathrm{F}}, \\
    m^*_\mathrm{L}&=k_{\mathrm{F}_y}\left.\left(\frac{\partial E(\bm k)}{\partial k_y}\right)^{-1}
    \right\rvert_{\bm k = \bm k_\mathrm{F}}.
\label{eq:effective_mass_formula}
\end{aligned}
\end{equation}
One can use the renormalized quasiparticle energy $E(\bm k)$ obtained by directly solving the self-consistent Dyson's equation [Eq.~(\ref{eq:Dyson})]. This way of calculating the effective mass is called the ``off-shell approximation''. It is obvious that the off-shell approximation provides the exact effective mass if the exact self-energy is used. Within the leading order dynamical screening RPA theory, however, the off-shell approximation mixes up perturbative orders in an inconsistent manner, and thus has been argued to be an inappropriate approximation to be used when we work with the leading order self-energy \cite{DuBois1959,DuBois1959a,Rice1965,Lee1975,Ting1975,Zhang2005,Zhang2005a}. For this reason, we use the on-shell effective mass approximation where we take only the first iterative solution of Dyson's equation, giving 

\begin{equation}
    E(\bm k) = \xi_{\bm k} + \mathrm{Re}\{\Sigma[\bm k,\xi_{\bm k}]\}.
    \label{eq:Dyson_equation_onshell}
\end{equation}
By putting Eq.~(\ref{eq:Dyson_equation_onshell}) into Eq.~(\ref{eq:effective_mass_formula}), we can obtain the expression for the on-shell effective mass:
\begin{equation}
\begin{aligned}
    m^*_\mathrm{H}&=\left\{\left. 1 + \frac{m_\mathrm{H}}{k_x}\frac{\partial\mathrm{Re}[\Sigma(\bm k,\xi_{\bm k})] }{\partial k_x}\right\rvert_{\bm k= \bm k_\mathrm{F}} \right\}^{-1}, \\
    m^*_\mathrm{L}&=\left\{\left. 1 + \frac{m_\mathrm{L}}{k_y}\frac{\partial\mathrm{Re}[\Sigma(\bm k,\xi_{\bm k})] }{\partial k_y}\right\rvert_{\bm k= \bm k_\mathrm{F}} \right\}^{-1}.
\label{eq:effective_mass_formula_final}
\end{aligned}
\end{equation}

\begin{figure}[!htb]
    \centering
    \includegraphics[width=0.9\linewidth]{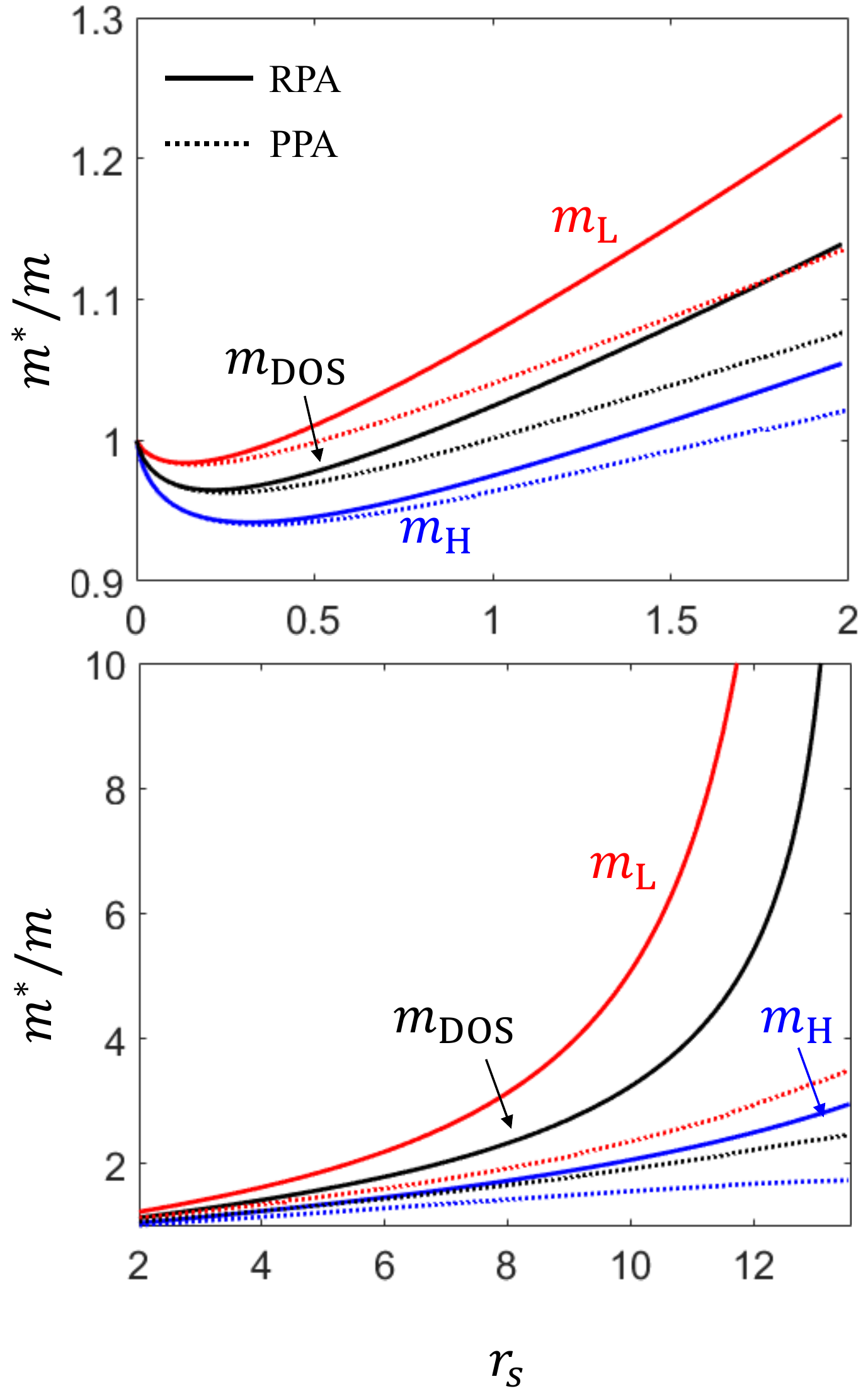}
    \caption{Calculated on-shell effective masses along the heavy-mass (red) and light-mass (blue) directions along with the calculated renormalized density-of-states mass $m^*_\mathrm{DOS}=\sqrt{m^*_\mathrm{H}m^*_\mathrm{L}}$. The solid (dashed) line represents effective masses calculated within the RPA (PPA). Here we set $m_\mathrm{H}/m_\mathrm{L}=10$.     }
    \label{fig:effective_mass}
\end{figure}

\begin{figure}[!htb]
    \centering
    \includegraphics[width=0.9\linewidth]{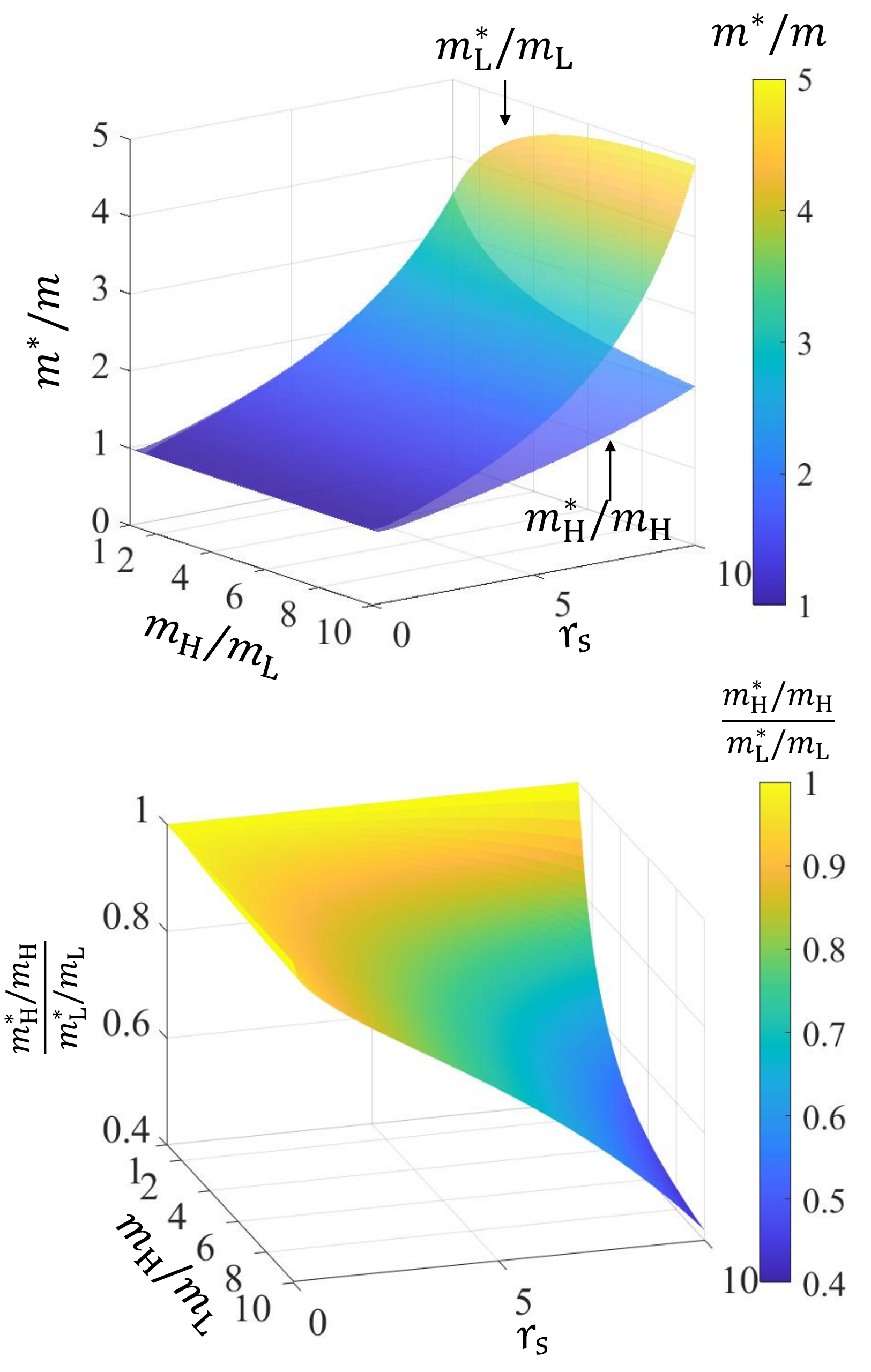}
    \caption{Three-dimensional plots of calculated on-shell effective masses (top) and the anisotropy of renormalization effective mass (bottom) as a function of $m_\mathrm{H}/m_\mathrm{L}$ and $r_s$. Note that the bare mass anisotropy is more suppressed with increasing $r_s$. }
    \label{fig:effective_mass2}
\end{figure}

Figure~\ref{fig:effective_mass} presents calculated effective masses $m^*_\mathrm{H}$ and $m^*_\mathrm{L}$ as a function of $r_s$ along with the renormalized density-of-states mass $m^*_\mathrm{DOS}=\sqrt{m^*_\mathrm{H}m^*_\mathrm{L}}$ obtained through multiplying the two calculated effective masses. It should be noted that the many-body mass enhancement is larger for $m_\mathrm{L}$ than $m_\mathrm{H}$, i.e., $m^*_\mathrm{L}/m_\mathrm{L}>m^*_\mathrm{H}/m_\mathrm{H}$, indicating the anisotropy of the system is reduced by interactions. It is also worth noting that $m^*_\mathrm{L}$ exhibits a divergent behavior at large $r_s\sim12$ whereas $m^*_\mathrm{H}$ shows relatively slowly increasing behavior at the equivalent $r_s$. Such a divergent behavior of the effective mass has been studied in depth theoretically in the corresponding isotropic 2D system \cite{Zhang2005}. Our results show that in the anisotropic system this effective mass divergence appears anisotropically with $m_\mathrm{L}$ ($m_\mathrm{H}$) having a smaller (larger) critical $r_s$ for this divergence. This is again a clear prediction of the theory which can be checked in anisotropic 2D systems, such as Si 110 and 111 inversion layers as well as 2D AlAs quantum wells and oxide heterostructures. Our results of course apply to any 2D electron or hole system with an elliptical bare band Fermi surface.
We mention that this theoretically predicted on-shell effective mass divergence is closely connected with the dispersion instability in strongly interacting fermionic systems \cite{Zhang2005b, Philippe1992, KHODEL1992, galitski2003}.
Our paper, therefore, indicates that the dispersion instability in anisotropic systems will be anisotropic with the instability happening first (i.e., at higher density) for the lower effective mass direction.  This is a non-obvious prediction, but given that the critical $r_s$ for the transition is much larger than unity, our RPA theory may not have quantitative accuracy in the regime of the transition.

In Fig.~\ref{fig:effective_mass2}, we show the calculated effective-mass renormalization in three-dimensional plots, clearly demonstrating how interaction suppresses the bare anisotropy by enhancing $m_\mathrm{L}$ and suppressing $m_\mathrm{H}$, so that the renormalized $m^*_\mathrm{H}/m^*_\mathrm{L}$ is quantitatively smaller than the bare anisotropy $m_\mathrm{H}/m_\mathrm{L}$.  This many-body suppression of the bare anisotropy is stronger for larger $r_s$ where interaction effects are stronger.
We mention here that some earlier works in the literature, using less complete theories, have found, consistent with our results, that the interaction induced many-body renormalization may suppress the effective mass anisotropy \cite{Wu1995, Roldan2006, Tolsma2016}. 

The effective masses obtained within the PPA are represented by the dotted lines in Fig.~\ref{fig:effective_mass}. Note that the PPA effective masses for small $r_s<0.2$ are in good quantitative and qualitative agreement with the RPA results. With increasing $r_s$, however, the PPA effective mass rapidly deviates from the RPA effective mass and just slowly increases near the critical $r_s$ of the RPA effective mass. Thus the PPA should be used only in the high density limit for calculating the effective mass of the anisotropic system. Note, however, that the PPA results show that the many-body mass correction for $m_\mathrm{L}$ is larger than for $m_\mathrm{H}$, leading to a consistent qualitative conclusion with the RPA that the anisotropy of the system is reduced by interactions.

\section{Exchange-Correlation Potential}

\begin{figure}[!htb]
    \centering
    \includegraphics[width=0.9\linewidth]{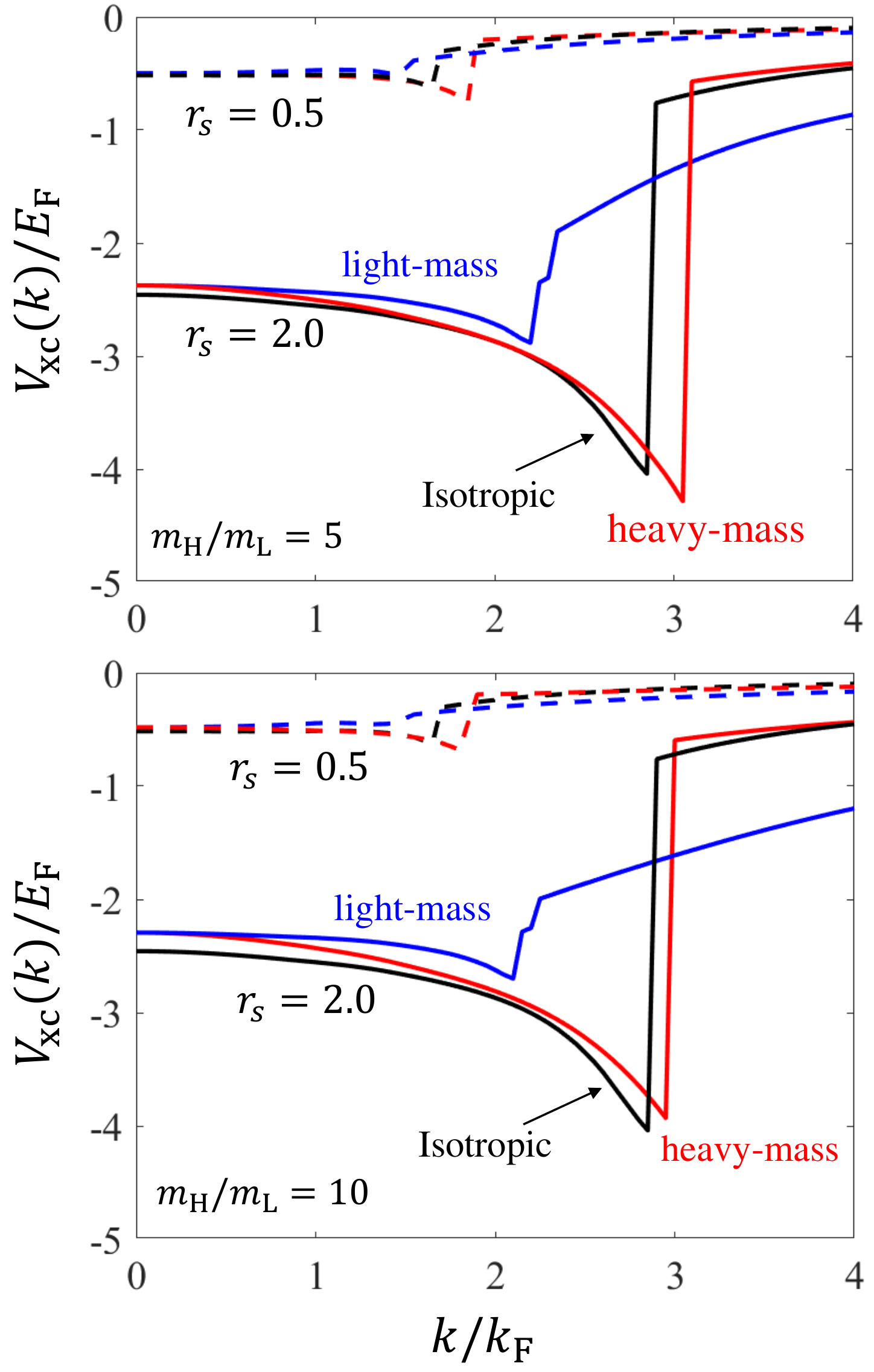}
    \caption{Calculated exchange-correlation function as a function of momentum $k$ along the heavy-mass (blue) and light-mass (red) directions for $r_s=0.5$ (dashed) and $2.0$ (solid), and $m_\mathrm{H}/m_\mathrm{L}=5$ (top) and $10$ (bottom). The black line represents the corresponding isotropic results with the density-of-states mass.  }
    \label{fig:exchance_correlation}
\end{figure}

A quantity of interest in the context of many-body theory is the exchange-correlation potential, $V_\mathrm{xc}(k)$, which provides a direct quantitative measure of the many-body renormalization effect:
\begin{equation}
V_\mathrm{xc} (k) = E(k) - \varepsilon_{\bm k}
\end{equation}
where $E(k)$ is the renormalized quasiparticle energy (i.e., the solution of the Dyson's equation [Eq.(\ref{eq:Dyson})]) and $\varepsilon_{\bm k}$ is the noninteracting bare energy defined by Eq.~(\ref{eq:noninteracting_energy}).  In Fig.~\ref{fig:exchance_correlation}, we show the calculated $V_\mathrm{xc}(k)$ within the full RPA theory for two values of $r_s$ and two values of mass anisotropy. We also show in this figure the corresponding results for the isotropic system using the density-of-states mass in each case. The sharp discontinuity in $V_\mathrm{xc}$ at a specific momentum is a real effect, arising from the sharp structures in the self-energy shown in Figs.~\ref{fig:self_energy_smallk} and \ref{fig:self_energy_largek}.  This sharp structure in $V_\mathrm{xc}$ is a direct manifestation of strong many-body effects in 2D.  For our specific purpose, it is interesting to note that $V_\mathrm{xc}$ manifests strong anisotropy, particularly at higher $r_s$ values where interaction effects are stronger.  


\section{Conclusion} \label{sec:conclusion}
We have provided a detailed investigation into the quasiparticle properties of a two-dimensional electron gas in the presence of mass anisotropy with an elliptic (instead of circular) noninteracting Fermi surface. Within the RPA theory we have studied mass anisotropy effects on the electron self-energy, the spectral function, the scattering rate, the exchange-correlation potential, and the renormalized effective mass.

We find that at small wavevectors the self-energies along the heavy-mass and light-mass directions are in good agreement near the quasiparticle solution, giving rise to almost identical quasiparticle peaks in the spectral functions along the two different directions. With increasing wavevector, the anisotropy of the self-energy becomes larger with a strong singular structure appearing in the self-energy along the light-mass direction, leading to a notable separation of the the quasiparticle peaks along different directions at large wavevectors ($k>2.0k_\mathrm{F}$). This result indicates that the renormalization effect on the quasiparticle state becomes more anisotropic as the quasiparticle energy increases from the Fermi surface. We compared our anisotropic results with the isotropic results calculated using the isotropic density-of-states mass, and find that the validity of the isotropic approximation is limited to only small wavevectors. Thus, for strong interactions, the quasiparticles are indeed anisotropic, but in general, the anisotropy is suppressed compared with the bare system.

Within the on-shell approximation, inelastic scattering rates along several different directions for various values of $r_s$ and mass ratios are calculated. We find that an abrupt rise in the scattering that occurs due to the quasiparticle decay via plasmon emissions shows up at different energies depending on the injected momentum direction because of the anisotropy of the plasmon dispersion. Although the scattering rate along the heavy-mass direction exhibits similar qualitative behaviors as the isotropic scattering rate, the scattering rate along a different direction, e.g., along the light-mass direction, shows a distinct behavior which has no isotropic analog. Thus, the inelastic scattering, which is experimentally accessible, manifests strong anisotropy in the interacting system. Our finding of qualitative anisotropic features in inelastic quasiparticle damping is an important result of the current paper, particularly since these anisotropic features cannot be simulated by any equivalent isotropic approximation.

Calculating the renormalized effective masses within the on-shell approximation, we show that, regardless of the value of $r_s$, the many-body mass enhancement is larger for $m_\mathrm{L}$ than for $m_\mathrm{H}$, implying that the bare anisotropy of the system is reduced by interactions.

Our calculated exchange-correlation potential manifests strong anisotropy, particularly for larger $r_s$ values.  A significant feature of our calculated $V_\mathrm{xc}$ is the presence of strong discontinuities at specific wave vectors, which arise from sharp structures in the 2D self-energy itself.

To investigate the validity of the plasmon-pole approximation for many-body calculations in the anisotropic system, we carried out calculations within the PPA for all the many-body quantities presented in this paper. We find that the PPA results in general are sufficiently close to the RPA results, qualitatively capturing most of the distinctive anisotropic features shown in our RPA results. There are, however, some quantitative inaccuracies in the plasmon-pole theory compared with the RPA results.

Our results on various many-body quantities show that the unjustified neglect of the mass anisotropy can result in an incorrect description of the quasiparticle properties of the anistropic system, missing interesting features originating from the mass anisotropy that cannot be captured even qualitatively by the commonly used approximation of using the isotropic density-of-states mass. On the other hand it is indeed true that the mass anisotropy itself is suppressed by interactions, and the renormalized mass ratio $m^*_\mathrm{H}/m^*_\mathrm{L}$ is always smaller than the bare mass ratio $m_\mathrm{H}/m_\mathrm{L}$, providing a weak justification for the extensively used isotropic quasiparticle approximation in interacting anisotropic systems.  In fact, the anisotropy suppression increases with increasing interaction. The reason for this interaction induced anisotropy suppression is that Coulomb interaction is always isotropic with perfect spherical symmetry, and in the interaction-dominated regime, anisotropy is consequently suppressed. There are, however, significant anisotropic features in the quasiparticle spectral function away from the Fermi surface and in the inelastic scattering rate at higher energies which cannot be captured by the isotropic approximation at all.

\acknowledgments
We thank A. Principi and M. Shayegan for communications on problems with citations in the first version of the paper. This work is supported by the Laboratory for Physical Sciences.

\appendix

\section{Fermi surface renormalization}\label{sec:appendixA}
We provide some results on the interaction induced Fermi surface topology modification, which has recently been discussed in Ref.~\cite{Ahn2020}.  The important point is that such Fermi surface shape modification is quantitatively very small, providing a justification for its neglect.  We emphasize that there is no symmetry reason to expect that the interacting Fermi surface should remain elliptic just because the noninteracting Fermi surface is elliptical since the spherical symmetry no longer applies in the presence of mass anisotropy.  We, however, find that the interacting system remains elliptic to a very high degree of accuracy, most likely because the Coulomb interaction itself is always isotropic and spherically symmetric.

\begin{figure}[htb]
    \centering
    \includegraphics[width=0.9\linewidth]{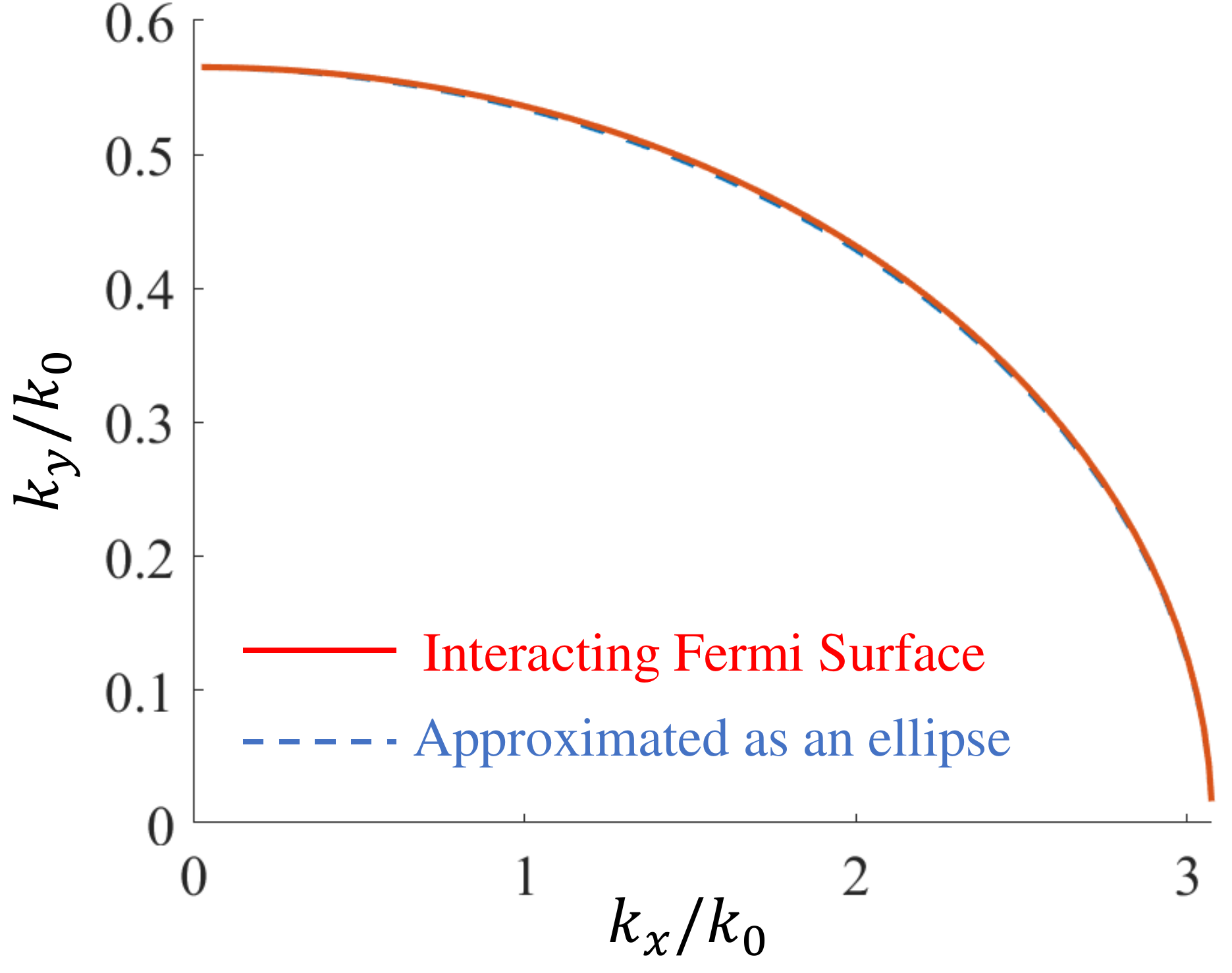}
    \caption{Non-elliptical interacting Fermi surface (red solid line) along with the interacting Fermi surface approximated as an ellipse (blue dashed line). Here we set $r_s=0.5$ and $m_\mathrm{H}/m_\mathrm{L}=10$.}
    \label{fig:Fermi_Surface}
\end{figure}

In Fig.~\ref{fig:Fermi_Surface} we show our calculated, for the full RPA self-energy theory, Fermi surface shape change for $r_s=0.5$ and a bare mass ratio of 10.  The shape change is minuscule, well below 0.1$\%$ in general, which is not of any experimental significance.  For larger $r_s$, the shape change is somewhat larger, but still never more than 0.1$\%$.  

\begin{figure}[htb]
    \centering
    \includegraphics[width=0.9\linewidth]{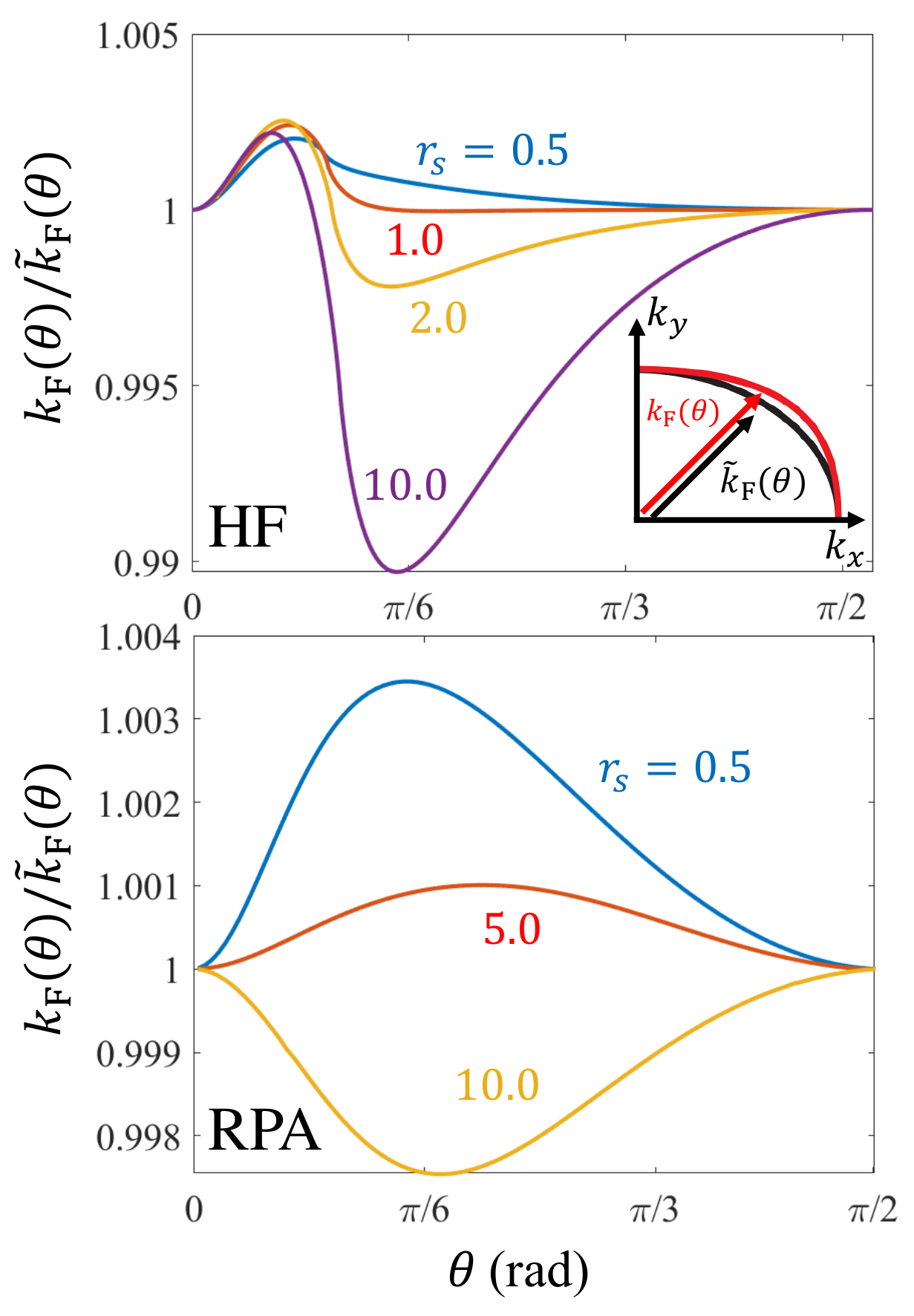}
    \caption{The ratio of the Fermi wavevector of the interacting Fermi surface [$k_F(\theta)$ in the inset] to that of the interacting Fermi surface approximated as an ellipse [$\widetilde{k}_F(\theta)$ in the inset] calculated within the Hartree-Fock approximation (top) and the RPA (bottom)}
    \label{fig:Fermi_Surface_HF_GW}
\end{figure}

In Fig.~\ref{fig:Fermi_Surface_HF_GW} we show the calculated Fermi surface shape change in the Hartree-Fock theory with just the exchange-only self-energy as given in Eq.~(\ref{eq:self_energy_ex}). This is performed straightforwardly by calculating the interacting chemical potential (and hence the interacting Fermi surface) using Eq.~(\ref{eq:interacting_mu}).  For the sake of comparison we also show the Fermi surface shape using the full RPA self-energy.  The point to note is that the quantitative magnitude of the change in the Fermi surface topology is very small ($\sim0.1\%$) in either approximation, and the two theories agree well up to $r_s\sim1$.  For larger $r_s$, the correlation effects become important, and the RPA theory manifests some qualitatively new effect as discussed recently in Ref. \cite{Ahn2020}. Note that although the Hartree-Fock theory provides a qualitatively incorrect result for the effective mass for all $r_s$ values, it is a reasonable description for the Fermi surface shape because the Fermi surface derives directly from the renormalized chemical potential which is well-approximated by the exchange energy, at least for $r_s$ values which are not too large.

\begin{figure*}[!ht]
    \centering
    \includegraphics[width=1\linewidth]{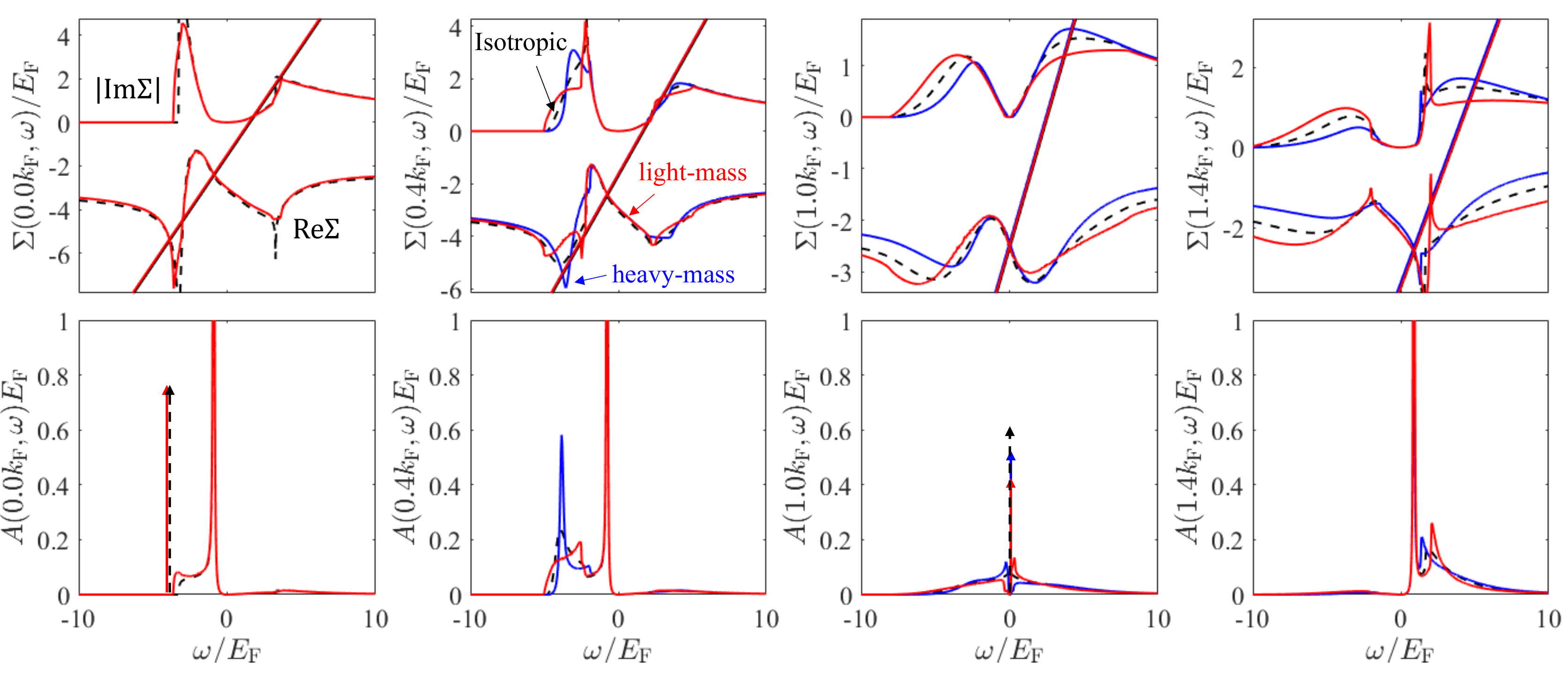}
    \caption{Calculated self-energies and the corresponding spectral functions for $k/k_\mathrm{F}=0.0$, $0.4$, $1.0$, and $1.4$. Here $m_\mathrm{H}/m_\mathrm{L}=5$ and $r_s=2.0$. The notation is the same as in Fig.~\ref{fig:self_energy_smallk}.}
    \label{app:fig:self_energy_smallk25}
\end{figure*}

\section{self-energy and spectral functions for different values of $r_s$ and the mass ratio} \label{sec:appendixB}
 In this Appendix, we show the Fig.~\ref{fig:self_energy_smallk} and Fig.~\ref{fig:self_energy_largek} results of the main text for larger $r_s$ as well as smaller mass ratio $m_\mathrm{H}/m_\mathrm{L}$ to demonstrate that the qualitative results of the self-energy and the corresponding spectral function in the main-text can be generalized, and to discuss quantitative effects of the two parameters ($r_s$ and $m_\mathrm{H}/m_\mathrm{L}$).

\begin{figure*}[!ht]
    \centering
    \includegraphics[width=1\linewidth]{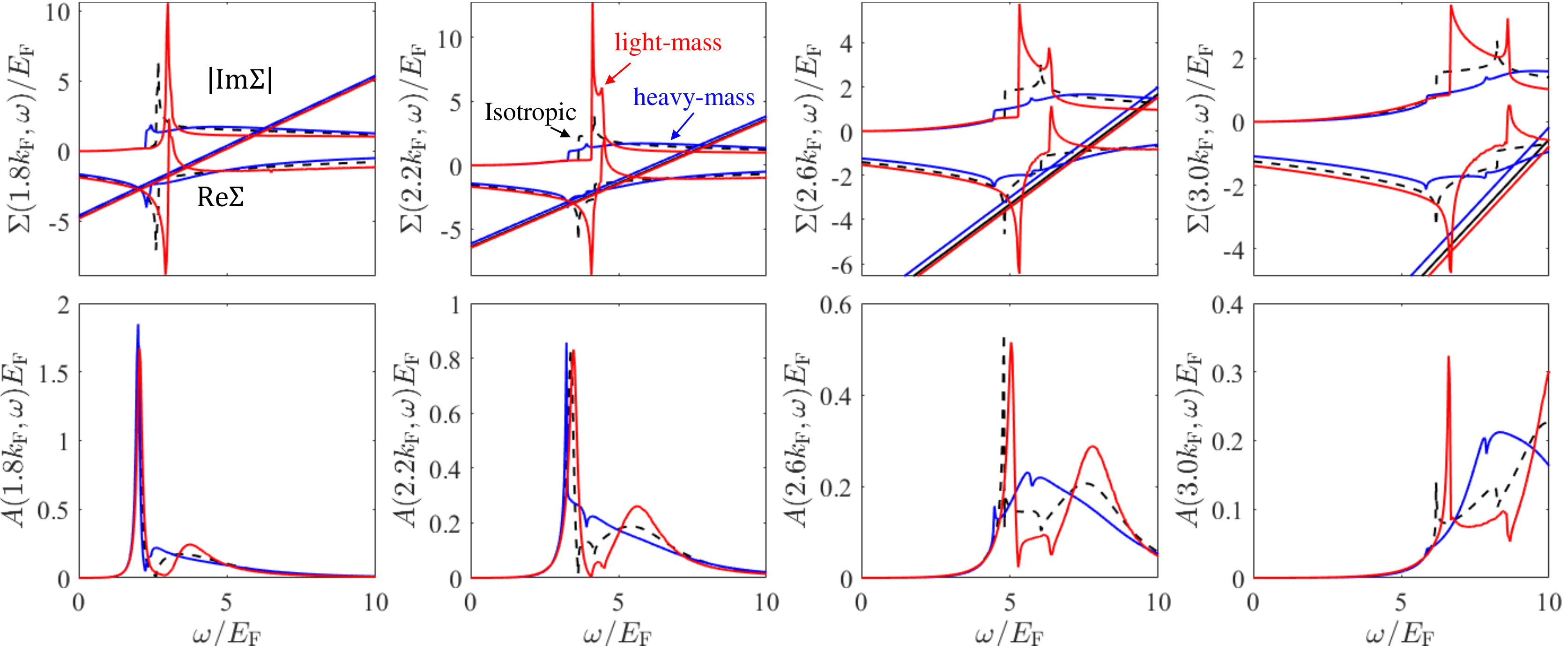}
    \caption{Calculated self-energies and the corresponding spectral functions larger wavevectors $k/k_\mathrm{F}=1.8$, $2.2$, $2.6$, and $3.0$. Here $m_\mathrm{H}/m_\mathrm{L}=5$ and $r_s=2.0$. The notation is the same as in Fig.~\ref{fig:self_energy_smallk} 
    }
    \label{app:fig:self_energy_largek25}
\end{figure*}
 
\subsection{Different mass ratios}
Figures~\ref{app:fig:self_energy_smallk25} and \ref{app:fig:self_energy_largek25} show calculated self-energies and the corresponding spectral functions for $m_\mathrm{H}/m_\mathrm{L}=5$ and $r_s=2.0$. Similarly as in the main-text results, for small wavevectors (Fig.~\ref{app:fig:self_energy_smallk25}), the quasiparticle peaks along the heavy- and light-mass directions are in good agreement and well approximated by the isotropic quasiparticle peak calculated using the density-of-states mass, while the plasmaron peak is anisotropic in the same way as the main-text results. The results for larger wavevectors (Fig.~\ref{app:fig:self_energy_largek25}) also show the same qualitative features as the main-text results with only a small quantitative difference that the separation of quasiparticle peaks along different directions is slightly smaller than that observed in the main-text results. 

\subsection{Different $r_s$'s}
\begin{figure*}[!htb]
    \centering
    \includegraphics[width=1\linewidth]{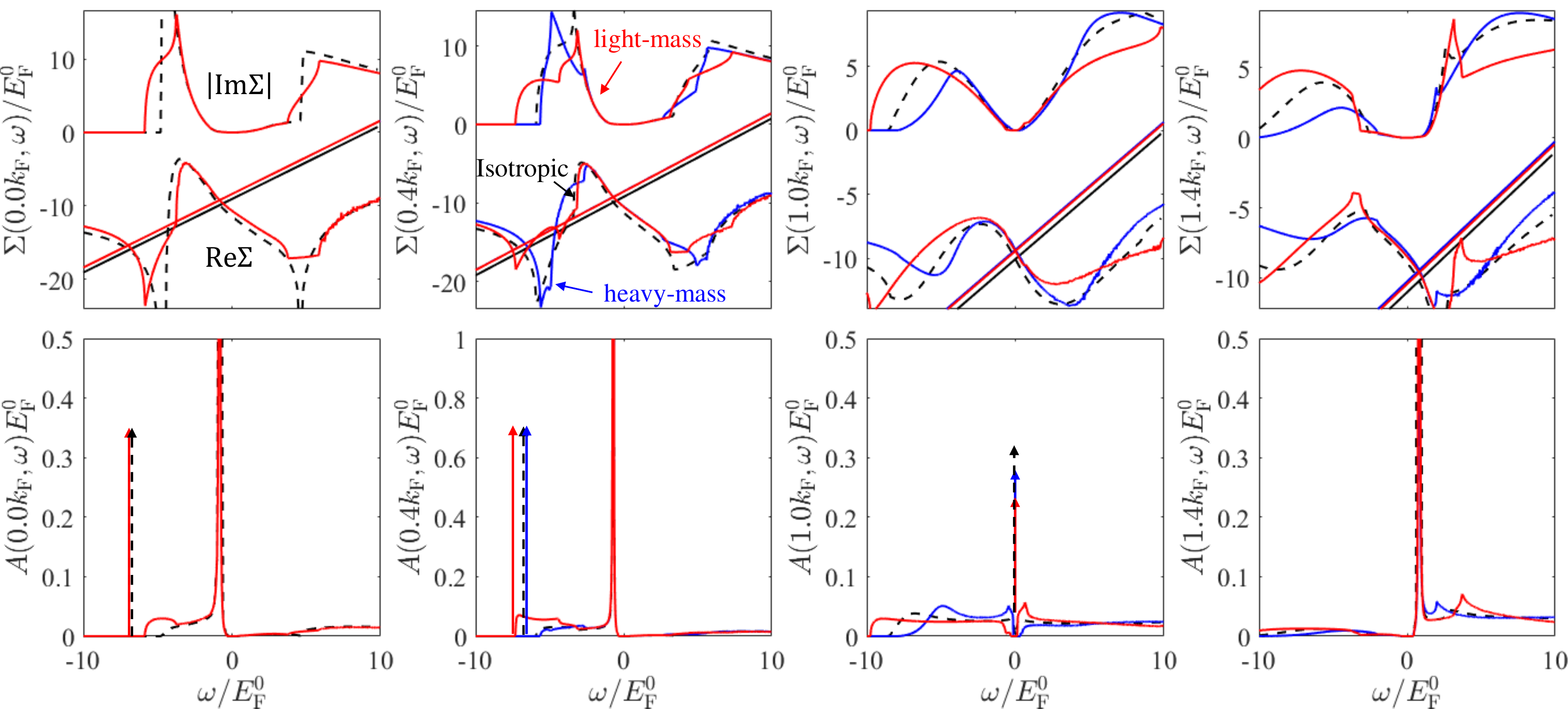}
    \caption{Calculated self-energies and the corresponding spectral functions for larger wavevectors $k/k_\mathrm{F}=0.0$, $0.4$, $1.0$, and $1.4$. Here $m_\mathrm{H}/m_\mathrm{L}=10$ and $r_s=6.0$. The notation is the same as in Fig.~\ref{fig:self_energy_smallk}}
    \label{app:fig:self_energy_smallk610}
\end{figure*}

\begin{figure*}[!htb]
    \centering
    \includegraphics[width=1\linewidth]{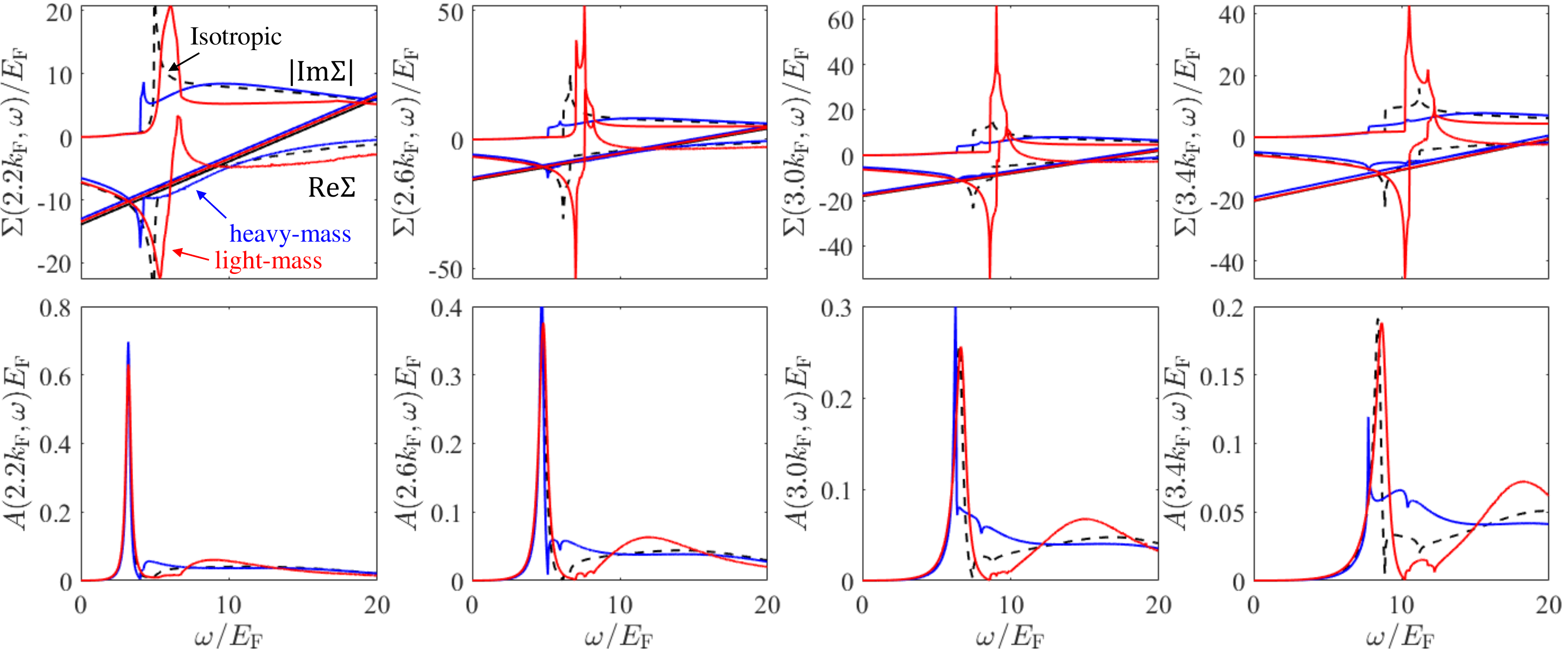}
    \caption{Calculated self-energies and the corresponding spectral functions for larger wavevectors $k/k_\mathrm{F}=2.2$, $2.6$, $3.0$, and $3.4$. Here $m_\mathrm{H}/m_\mathrm{L}=10$ and $r_s=6.0$. The notation is the same as in Fig.~\ref{fig:self_energy_smallk}}
    \label{app:fig:self_energy_largek610}
\end{figure*}

Figures~\ref{app:fig:self_energy_smallk610} and \ref{app:fig:self_energy_largek610} show calculated self-energies and the corresponding spectral functions for $m_\mathrm{H}/m_\mathrm{L}=10$ and $r_s=6.0$. For small wavevectors (Fig.~\ref{app:fig:self_energy_smallk610}), the quasiparticle peaks along the heavy- and light-mass directions are in good agreement similar as in the main-text results. Note, however, that the plasmaron peak for $k=0.4k_\mathrm{F}$ along the light-mass direction is undamped in contrary to the result for $r_s=2.0$ given in the main text where the corresponding plasmaron peak for $k=0.4k_\mathrm{F}$ is significantly damped (Fig.~\ref{fig:self_energy_smallk}). This indicates that the plasmaron peak along the light-mass direction for small wavevectors ($k<k_F$) is better defined with increasing $r_s$. For $k=0$ and $k=1.4k_\mathrm{F}$, the anisotropic plasmaron peaks are qualitatively similar to those in the corresponding main-text results.
For large wavevectors (Fig.~\ref{app:fig:self_energy_largek610}), the quasiparticle peaks along different directions are more separated with increasing $k$ similar as the main-text result but with a small quantitative difference that for the same wavevector the separation of the quasiparticle peaks is smaller compared to the main-text results.


\clearpage
\bibliography{ref}
\end{document}